\documentclass[11pt]{article}
\usepackage{graphicx}
\usepackage{longtable}
\usepackage[utf8]{inputenc}
\usepackage{geometry}
\usepackage{tabularx}
\usepackage{helvet} 
\usepackage{amsmath, amssymb}
\usepackage{hyperref}
\usepackage{url}
\usepackage{eurosym}
\usepackage{titlesec}
\usepackage{fancyhdr}
\usepackage{newunicodechar}
\newunicodechar{₂}{$_2$}
\usepackage{cite}
\usepackage{enumitem}
\usepackage{titling}
\usepackage{booktabs}
\usepackage{afterpage}
\usepackage{acronym} 
\usepackage{xcolor}
\usepackage{wrapfig}
\usepackage{setspace}
\usepackage{tikz}

\usetikzlibrary{shapes.geometric, arrows.meta, positioning, fit}

\definecolor{fortissblue}{RGB}{0,70,150}
\colorlet{dimSSE}{fortissblue}            
\colorlet{dimAI}{fortissblue!55!white}    
\colorlet{dimCON}{fortissblue!70!black}   
\colorlet{dimCROSS}{gray!65!black}   

\usepackage{caption}
\makeatletter
\long\def\@makefntext#1{%
  \tiny\parindent 1em\noindent
  \hb@xt@1.8em{\hss$\m@th^{\@thefnmark}$}#1}
\makeatother

\newcommand{\dimtag}[2]{\textcolor{#1}{\footnotesize$\blacksquare$}\,#2}
\titleformat{\section}{\color{fortissblue}\normalfont\Large\bfseries}{\thesection}{1em}{}
\titleformat{\subsection}{\color{fortissblue}\normalfont\large\bfseries}{\thesubsection}{1em}{}
\titleformat{\subsubsection}{\color{fortissblue}\normalfont\normalsize\bfseries}{\thesubsubsection}{1em}{}
\begin{document}
\pagenumbering{roman}  
\setcounter{page}{1}   

\begin{titlepage}
    \centering
    \vspace*{3cm}
    {\color{fortissblue}\Huge\bfseries Software-Defined Defence \\}
    \vspace{2cm}
    {\color{fortissblue}\LARGE A perspective on civilian-to-defence research transfer to SDD\par}
    \vspace{1cm}
    {\color{fortissblue}White Paper\par}

    \vfill
    \vspace{0.5cm}
    {\color{fortissblue}\small  Rute C. Sofia, Daniel Mendez, Simon Barner, Hao Shen, Julian Wörmann, Andrea Stocco, Axel von Arnim, Holger Pfeifer, Alexander Pretschner \\fortiss GmbH, Munich, Germany \par}
    \vspace{0.5cm}
    {\color{fortissblue}\large \today \par}
    \vfill
    \includegraphics[width=0.3\textwidth]{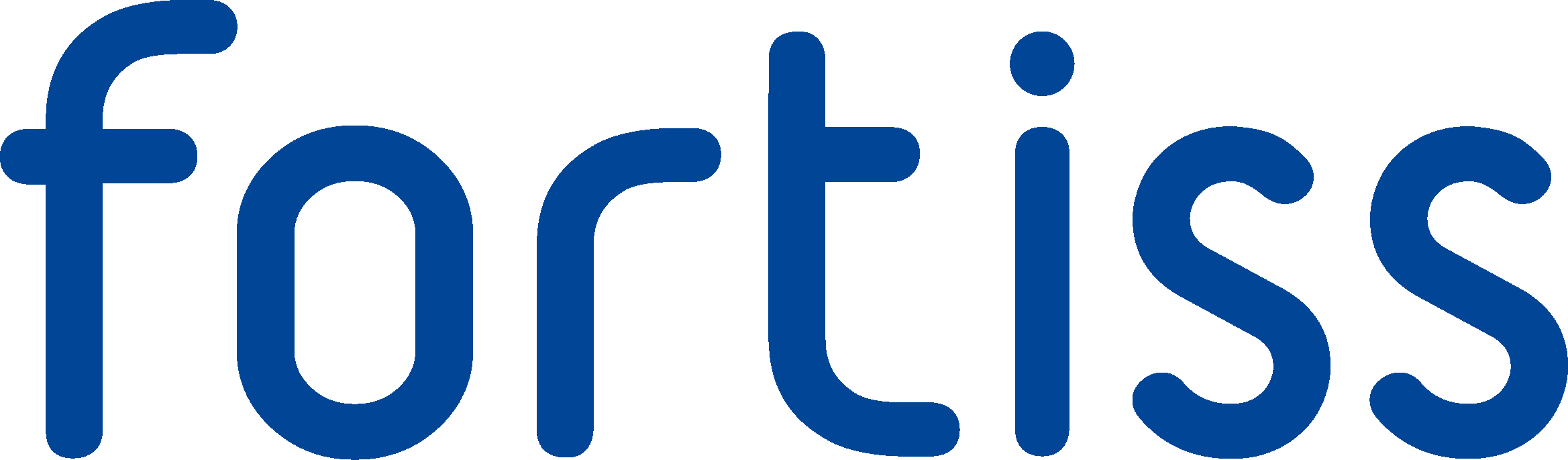} 
    
    \vspace*{2cm}
\end{titlepage}


\section*{Executive Summary}
Military capability is increasingly determined by software. However, defence platforms are procured on a decade-long timescale, while the software and Artificial Intelligence (AI) models they carry must evolve in days or hours. 

This white paper calls this mismatch the lifecycle paradox, and argues that it is the central engineering problem \textit{Software-Defined Defence (SDD)} must solve.

The central thesis of this paper is that SDD is based on three intertwined dimensions: \textbf{software and systems engineering} (how platforms are designed, procured, and certified), \textbf{AI engineering} (whether learned and autonomous components can be sovereign and trusted), and \textbf{connectivity \& infrastructure engineering} (whether sensors, AI, and operators can actually exchange information in time). 

The proposed way to reach resilient SDD, considering as starting point not a clean slate, but technologies developed for civilian purposes, is to address capabilities required to SDD  based on a continuous, DevOps-style engineering loop: \textbf{model-based systems engineering and simulation-based testing front-load design and verification}; \textbf{tactical connectivity and low-power edge execution} carry that design into contested operation; and \textbf{continuous cybersecurity compliance}, \textbf{continuous assurance}, and \textbf{variability management} run beneath both phases as cross-cutting concerns.

Such a DevOps SDD loop is sustainable, considering existing capabilities and methodologies that have been built and proven in domains such as Automotive, Manufacturing, Space, Energy. The paper advocates that now it is important to take steps that allow the validation of such capabilities under genuinely adversarial, contested, or defence-certified conditions. To do so, specific representative capabilities and assets are described, explaining existing gaps and how to address such gaps from a short, medium, and long-term (2030) perspective.

Closing the SDD gap in a way that meets European and German legislation (e.g., AI Act, GDPR) while keeping civic benefits is a distributed responsibility: researchers must redirect existing methods toward adversarial conditions; industry must expose civilian tooling to real operational requirements; policymakers must shape the regulatory instruments the transfer depends on; and defence agencies must validate the result with real operators. Recommendations on how to go ahead close the paper across three horizons. First, a short-term baseline (2026–2027) proposes to establish adversarial testing, regulatory mapping, and jamming-resilient connectivity pilots. Second, a medium-term pipeline (2027–2029) must be developed, demonstrating incremental certification, an open robustness benchmark, and tactical end-to-end orchestration. Thirdly, a long-term validation (2029–2030) closing the full loop under real operational and human-machine-teaming conditions has to be pursued.

\textbf{fortiss positions itself as a technical partner across all three SDD dimensions} offering open-source tooling, ongoing standards contributions, and a research team ready to build demonstrators from defined use cases. We invite defence stakeholders, research institutions, and policymakers to jointly turn this evidence base into defence-ready capability.

\vspace*{0.5cm}
\textbf{\color{fortissblue}Keywords:} Software-Defined Defence; AI engineering; systems engineering; tactical connectivity; low-power edge intelligence, autonomous systems; cybersecurity compliance.

\newpage
\tableofcontents
\addcontentsline{toc}{section}{Executive Summary}
\newpage

\newpage
\listoffigures
\addcontentsline{toc}{section}{List of Figures}
\listoftables
\addcontentsline{toc}{section}{List of Tables}
\newpage

\newpage

\section*{List of Acronyms}

\begin{longtable}{@{}p{3cm}p{12.5cm}@{}}
\textbf{5G} & Fifth Generation Mobile Communication \\
\textbf{6G} & Sixth Generation Mobile Communication \\
\textbf{ADNS} & Automated Digital Network System \\
\textbf{ADS} & Automated Driving System(s) \\
\textbf{AF3} & AutoFOCUS3 \\
\textbf{AI} & Artificial Intelligence \\
\textbf{AP} & Access Point \\
\textbf{B5G} & Beyond 5G \\
\textbf{BDLI} & Bundesverband der Deutschen Luft- und Raumfahrtindustrie (Federal Association of the German Aerospace Industry) \\
\textbf{BDSV} & Bundesverband der Deutschen Sicherheits- und Verteidigungsindustrie (Federal Association of the German Security and Defence Industry) \\
\textbf{Bitkom} & Bundesverband Informationswirtschaft, Telekommunikation und neue Medien (Federal Association for Information Technology, Telecommunications and New Media) \\
\textbf{BMVg} & Bundesministerium der Verteidigung (German Federal Ministry of Defence) \\
\textbf{C2} & Command and Control \\
\textbf{CPS} & Cyber-Physical Systems \\
\textbf{CRA} & Cyber Resilience Act \\
\textbf{DARPA} & Defense Advanced Research Projects Agency \\
\textbf{DDIL} & Denied, Disrupted, Intermittent, and Limited \\
\textbf{DDS} & Data Distribution Service \\
\textbf{DetNet} & Deterministic Networking \\
\textbf{DTN} & Delay/Disruption-Tolerant Networking \\
\textbf{DTNMA} & DTN Management Architecture \\
\textbf{EDF} & European Defence Fund \\
\textbf{ERCC} & Emergency Response Coordination Centre \\
\textbf{ESA} & European Space Agency \\
\textbf{ETSI} & European Telecommunications Standards Institute \\
\textbf{EU} & European Union \\
\textbf{FMEA} & Failure Mode and Effects Analysis \\
\textbf{FMN} & Federated Mission Networking \\
\textbf{FTA} & Fault Tree Analysis \\
\textbf{FTM} & Fine Time Measurement \\
\textbf{GenAI} & Generative Artificial Intelligence \\
\textbf{GPS} & Global Positioning System \\
\textbf{HARA} & Hazard Analysis and Risk Assessment \\
\textbf{ICN} & Information-Centric Networking \\
\textbf{IEC} & International Electrotechnical Commission \\
\textbf{IESG} & Internet Engineering Steering Group \\
\textbf{IETF} & Internet Engineering Task Force \\
\textbf{IHL} & International Humanitarian Law \\
\textbf{IoT} & Internet of Things \\
\textbf{IP} & Internet Protocol \\
\textbf{IRTF} & Internet Research Task Force \\
\textbf{ISO} & International Organization for Standardization \\
\textbf{ISR} & Intelligence, Surveillance, and Reconnaissance \\
\textbf{LiDAR} & Light Detection and Ranging \\
\textbf{LLM} & Large Language Model \\
\textbf{MAC} & Medium Access Control \\
\textbf{MBSE} & Model-Based Systems Engineering \\
\textbf{MDO} & Multi-Domain Operations \\
\textbf{ML} & Machine Learning \\
\textbf{MOSA} & Modular Open Systems Architecture \\
\textbf{MQTT} & Message Queuing Telemetry Transport \\
\textbf{NATO} & North Atlantic Treaty Organization \\
\textbf{NDN} & Named Data Networking \\
\textbf{OFDMA} & Orthogonal Frequency-Division Multiple Access \\
\textbf{OWASP} & Open Worldwide Application Security Project \\
\textbf{PHY} & Physical Layer \\
\textbf{QoS} & Quality of Service \\
\textbf{RAW} & Reliable and Available Wireless \\
\textbf{SAE} & SAE International \\
\textbf{SBOM} & Software Bill of Materials \\
\textbf{SDD} & Software-Defined Defence \\
\textbf{SDN} & Software-Defined Networking \\
\textbf{SDO} & Standards Development Organization(s) \\
\textbf{SDR} & Software-Defined Radio \\
\textbf{Sim2Real} & Simulation-to-Reality \\
\textbf{SME} & Small and Medium Enterprise(s) \\
\textbf{SPES} & Software Platform Embedded Systems \\
\textbf{STPA} & System-Theoretic Process Analysis \\
\textbf{SyNAPSE} & Systems of Neuromorphic Adaptive Plastic Scalable Electronics \\
\textbf{SysML} & Systems modelling Language \\
\textbf{TARA} & Threat Analysis and Risk Assessment \\
\textbf{TAS} & Time Aware Shaper \\
\textbf{TRL} & Technology Readiness Level \\
\textbf{TSN} & Time Sensitive Networking \\
\textbf{TWT} & Target Wake Time \\
\textbf{UAS} & Unmanned Aircraft System(s) \\
\textbf{V\&V} & Verification and Validation \\
\textbf{VNF} & Virtualised Network Function \\
\textbf{WIN-T} & Warfighter Information Network -- Tactical \\
\end{longtable}

\newpage

\clearpage
\pagenumbering{arabic}  
\setcounter{page}{1}   

\section{Introduction: The Software-Defined Turn in Defence}
\label{sec:introduction}

Software has become the primary determinant of military capability. Defence platforms, once defined by their hardware, increasingly depend on autonomous software that fuses sensor data, coordinates autonomous behaviour, adapts to contested environments, and requires updates on a timescale that hardware procurement was never designed to match. Recent conflicts have already produced concrete operational lessons that confirm this shift, and Germany, Europe, and NATO have responded with matching policy and funding measures.

This white paper sets out the fortiss research perspective on \emph{Software-Defined Defence} (SDD), an engineering challenge spanning three necessary and complementary dimensions: a \textbf{software and systems engineering} dimension concerned with how defence platforms are designed, procured, and certified; an \textbf{\textit{Artificial Intelligence (AI)} engineering} dimension concerned with the trustworthiness of the learned and autonomous components those platforms increasingly rely on; and a \textbf{connectivity and infrastructure engineering} dimension concerned with whether the underlying network and computational substrate can deliver information between sensors, AI components, and operators within operationally meaningful bounds. Treating any one of these dimensions in isolation understates what SDD actually requires.

Much of the engineering substance behind these three dimensions has a well-established scientific basis in specific areas of computer science, developed independently of any defence application. The software and systems engineering dimension draws on model-based systems engineering, formal verification, and software architecture research; the AI engineering dimension draws on \textit{Machine Learning (ML)} robustness, uncertainty quantification, and explainable AI; and the connectivity and infrastructure engineering dimension draws on computer networking, distributed systems, and edge-cloud computing. Methods from these areas have been developed and validated in automotive, manufacturing, energy, civil-infrastructure settings, often under dependability requirements as demanding as those defence platforms must meet. \textbf{Civil-to-defence technology adaptation} is the core thesis of this white paper. The central engineering question is not whether such civilian-proven technology is relevant to SDD, but what specifically must be added, tested, or hardened to make it trustworthy under contested, adversarial, and denied conditions that civilian development was never designed to anticipate. This same technology base is directly relevant to hybrid-warfare resilience. The attack surface of a critical energy or communications network under sustained cyber and disinformation pressure already resembles that of a defence C2 system, and the capabilities this paper sets out apply to both without modification.

To bring the SDD vision into reality by 2030, this paper makes four contributions, considering a broad audience composed of researchers, industry, defence agencies, civil society, and policymakers. First, it frames SDD as a interconnected set of engineering challenges that span three necessary and complementary dimensions, software and systems engineering, AI engineering, and connectivity and infrastructure engineering, and names the \textbf{lifecycle paradox}, the mismatch between decade-long hardware cycles and software that must change on a timescale of days, as the structural tension running across all three. Second, it turns this tension into a continuous, DevOps-inspired capability loop covering design, verification, deployment, and operation. Third, it assesses this loop against illustrative fortiss and related research, stating explicitly, capability by capability, what is already validated in civilian and dual-use settings and what remains an open transfer problem under contested, adversarial, or defence-certified conditions. Fourth, it turns this assessment into concrete, time-horizoned actions for researchers, industry, policymakers, and defence agencies.

The remainder of this paper is organized as follows. \autoref{sec:SDD-definitions} defines SDD and positions it within the broader shift from cyber-physical systems to software-defined systems, introduces the geopolitical and regulatory pressures driving the shift, and names the \emph{lifecycle paradox} that runs through the rest of the paper. \autoref{challenges} details the engineering challenges that this paradox creates, from certification to tactical connectivity. \autoref{sec:capabilities} sets out proposed capabilities for addressing these challenges. \autoref{sec:contributions} positions concrete fortiss research concepts and assets against these capability areas and is explicit about where each asset is proven and where it remains an untested transfer case. \autoref{sec:vision} sets out what researchers, industry, policymakers, and defence agencies can do to close the gap, organised by time horizon and by the three dimensions of the fortiss SDD vision.

\section{From Cyber-Physical Systems to Software-Defined Defence}
\label{sec:SDD-definitions}
Defence platforms are a specialised class of \textit{Cyber-Physical Systems (CPS)}: software-driven systems that sense, reason about, and act on their physical environment. This section traces how that CPS foundation is evolving into SDD, starting with its roots in CPS engineering (\autoref{sec:cps-sdd}), moving to the geopolitical and investment pressures now accelerating it (\autoref{sec:geopolitical}), and arriving at a definition of SDD and its three constituent dimensions (\autoref{sec:SDD-dimensions}), the basis for the fortiss SDD vision. \autoref{sec:convergence} then brings these dimensions together, setting out the four factors converging to make SDD an operational necessity rather than an engineering option. These subsections motivate what this paper names the \textbf{lifecycle paradox}, a structural mismatch between two timescales: defence hardware is procured over decades, while the software it carries must be updated within days to hours.

\subsection{From CPS to Software-Defined Systems}
\label{sec:cps-sdd}
Software and systems engineering has treated hardware as a constraint rather than a starting point for decades. The system landscape and the engineering challenges that such a landscape needs to overcome have evolved continuously and rapidly ever since. CPS in particular, have drawn significant attention in recent years: software-driven systems that interact with the physical world through sensors, actuators, and networked computation, operating in continuous feedback loops in which physical processes affect computation and vice versa~\cite{lee2008cps,rajkumar2010cps}. Germany has formulated this paradigm shift, from classical embedded control to globally networked software-intensive systems, through its \textit{National Academy for Science and Engineering (acatech)}\footnote{https://en.acatech.de/} agenda on CPS~\cite{broy2010cps_acatech,
geisberger2012agendaCPS}. This agenda identifies three structural characteristics that distinguish modern CPS from earlier generations of embedded systems~\cite{dagstuhl2019es4cps,
geisberger2012cps_vision}. First, CPS perform \textbf{complex real-time computation}: computation proceeds under non-negotiable timing, resource, and safety constraints, so that correctness is inseparable from timeliness. Second, CPS exhibit \textbf{hybrid discrete-continuous control}, processing both event-driven logical transitions and continuous signal dynamics, and coupling software state machines with physical plant models. Third, CPS depend on \textbf{distributed, networked communication}, in which components, heterogeneous and frequently mobile, interact across contested and bandwidth-variable channels.

Scale, heterogeneity, adaptive connectivity, and cognition set modern CPS apart from earlier generations. \textit{Internet of Things (IoT)}, Digital twins, edge AI, swarm coordination, and 5G/6G integration are expanding what CPS can sense, reason about, and act at operational speed~\cite{broy2012cps_position,
geisberger2012agendaCPS}.

\textbf{Software-defined CPS} carry this logic further. \textit{Software-Defined Networking (SDN}) decoupled the control plane from the data plane and put it into programmable software~\cite{kathiravelu2017sdcps}. Applying the same principle to CPS produces a fundamental architectural reorientation: mechanical and structural platform elements may remain stable for decades, and electronic, compute, sensing, and communication subsystems refresh on multi-year cycles, but the differentiating capabilities are increasingly defined, extended, and updated through software, AI models, and configuration. Five properties follow:

\begin{itemize}
  \item \textbf{Reconfigurability}: mission profiles, sensor fusion
    algorithms, communication waveforms, and rule-of-engagement
 enforcement logic can change without hardware modification.
  \item \textbf{Modularity}: clear interfaces allow software
    components to be independently developed, validated, and reused
    across platform families~\cite{broy2013engineering_cps}.
  \item \textbf{Scalability}: new functional modules are deployed
    against stable platform APIs, an app-store model for embedded
    systems.
  \item \textbf{Resilience through self-adaptation}: runtime
    monitoring enables graceful degradation and dynamic
    risk-acceptance adjustment rather than failure
 behaviour~\cite{fhg_iks2026sdd}.
  \item \textbf{Lifecycle decoupling}: hardware procurement cycles
    (decade-scale) are structurally separated from software evolution
    (week-to-month cadence for patches, hours for AI model
    updates)~\cite{bmvg2023sdd}.
\end{itemize}

This last property has the most profound implications for defence, and this paper calls it the \textbf{lifecycle paradox}. \autoref{sec:geopolitical} turns to the geopolitical and investment pressures that have made this paradox urgent.

\subsection{The Geopolitical Inflection Point}
\label{sec:geopolitical}
European defence procurement has until recently operated under the assumption that system acquisition cycles of 10 to 20 years were structurally acceptable. Now, two converging pressure points challenge this assumption. The first pressure point concerns operational evidence from contemporary systems. Field experience from the conflict in Ukraine shows that the capacity to rapidly iterate software, integrate commercial AI and communication capabilities, and update autonomous system behaviour can influence battlefield outcomes on a timescale of months, rather than years. AI-enabled navigation for \textit{Unmanned Autonomous Systems (UAS)} systems is a concrete example, associated with substantial improvements in mission effectiveness~\cite{ec2025roadmap}. The resilience of critical infrastructures including telecommunications and healthcare infrastructures to cybersecurity offers a second distinct example. Continuously evolving attack surfaces and patterns, often leading to broken access control, security misconfiguration, or software supply chain and authentication failures, account for most of the top ten security risks identified by OWASP\footnote{https://owasp.org/Top10/2025}. Addressing risks like these at speed is only possible through automation: pattern recognition, automated reasoning, adaptive decision support. Hence, the agility of the software layer has become an operationally relevant factor in its own right\footnote{https://corvusintell.com/blog/defense-market/defense-tech-market-europe-2025/}.

The second pressure point arises from a shift in defence investment commitments across NATO member states and across Europe. The number of NATO members meeting the 2\% GDP spending target rose sharply between 2020 and 2024, reaching 23 of 32 members in 2024~\footnote{https://www.cfr.org/expert-brief/nato-countries-reach-spending-milestone-2-percent-enough}. Several members have committed to higher targets still: Poland at 4\%, the Baltic states between 3 and 3.5\%. Germany's \textit{Sondervermögen} of €100 billion represents a structural commitment to higher defence investment that extends well beyond the fund itself~\footnote{https://www.bmvg.de/en/news/over-eur-100-billion-for-the-bundeswehr-and-for-our-security-5362626}. 

Furthermore, \textit{Command and control} (C2) software represents the largest single software segment in European defence. Current C2 modernization programs, including the German D-LBO,\footnote{https://www.bundeswehr.de/de/meldungen/digitalisierung-landbasierte-operationen}, and the French SCORPION\footnote{https://www.arquus-defense.com/scorpion-program}, are mainly focused on replacing proprietary, monolithic C2 stacks with software-defined architectures that can be updated independently of the underlying hardware. The main share of funding and engineering effort in these programmes lies therefore in \textbf{software architecture adaptation}, as this is a must to support and adapt to different types of hardware.

As response, the EU has introduced a set of governance and financial directive defence instruments: the \emph{White Paper on European Defence - Readiness~2030}~\cite{ec2025whitepaper}; the \emph{EU Defence Industry Transformation Roadmap}~\cite{ec2025roadmap}; the SAFE loan instrument of \EUR{150}~billion\footnote{https://commission.europa.eu/topics/defence/future-european-defence}; and
the AGILE fast-fielding fund of \EUR{115}~million, targeting grant award within four months and technology deployment to armed forces within one to three years\footnote{https://commission.europa.eu/priorities-2024-2029/security-and-defence}. There is therefore a shift in EU policy from \textbf{reactive}, \textbf{fragmented} procurement towards \textbf{integrated}, \textbf{iterative} capability development. However, funding is not the same as engineering processes, as debated next.

\subsection{Three Dimensions of Software-Defined Defence}
\label{sec:SDD-dimensions}
The term SDD was formalised in the joint position paper of the \textit{Bundesministerium der Verteidigung} (BMVg, the German Federal Ministry of Defence), the \textit{Bundesverband der Deutschen Sicherheits- und Verteidigungsindustrie} (BDSV, the Federal Association of the German Security and Defence Industry), the \textit{Bundesverband der Deutschen Luft- und Raumfahrtindustrie} (BDLI, the Federal Association of the German Aerospace Industry), and the \textit{Bundesverband Informationswirtschaft, Telekommunikation und neue Medien} (Bitkom, the Federal Association for Information Technology, Telecommunications and New Media~\cite{bmvg2023sdd}). Its central premise: software determines the differentiating capabilities of defence systems, while the underlying mechanical platforms and electronic subsystems provide slower-evolving constraints and interfaces. The objective is to exploit software's capacity for continuous capability improvement, such as shorter development cycles, flexible capability insertion, scalability across platform variants, resilience through updateable and reconfigurable behaviour~\cite{bmvg2023sdd, geisberger2012agendaCPS}.

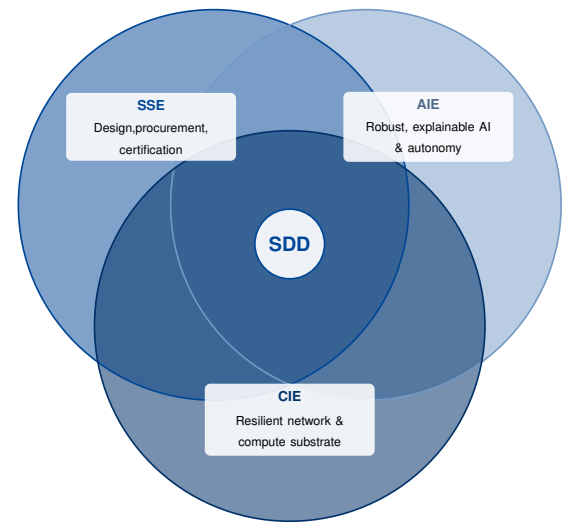
\begin{wrapfigure}{r}{0.42\textwidth} 
\centering
\resizebox{0.4\textwidth}{!}{%
\begin{tikzpicture}[font=\small]
  \coordinate (A) at (0,0);
  \coordinate (B) at (3.9,0);
  \coordinate (C) at (1.95,-3.1);
  \def\rad{5cm}
  \fill[dimSSE, opacity=0.50] (A) circle (\rad);
  \fill[dimAI,  opacity=0.50] (B) circle (\rad);
  \fill[dimCON, opacity=0.54] (C) circle (\rad);
  \draw[dimSSE, very thick] (A) circle (\rad);
  \draw[dimAI,  very thick] (B) circle (\rad);
  \draw[dimCON, very thick] (C) circle (\rad);
  \node[fill=white, fill opacity=0.88, text opacity=1, rounded corners,
        inner sep=5pt, align=center, text width=4cm] at (-1.6,2.0)
    {\textbf{\color{dimSSE}SSE}\\[2pt]
     \footnotesize\color{black} Design,procurement, \\certification};
  \node[fill=white, fill opacity=0.88, text opacity=1, rounded corners,
        inner sep=5pt, align=center, text width=4cm] at (5.5,2.0)
    {\textbf{\color{dimAI!70!black}AIE}\\[2pt]
     \footnotesize\color{black} Robust,~explainable AI \\ \& autonomy};
  \node[fill=white, fill opacity=0.88, text opacity=1, rounded corners,
        inner sep=5pt, align=center, text width=4cm] at (1.95,-5.5)
    {\textbf{\color{dimCON}CIE}\\[2pt]
     \footnotesize\color{black} Resilient network \& compute substrate};
  \node[fill=white, fill opacity=0.92, text opacity=1, draw=dimSSE,
        very thick, circle, inner sep=7pt,
        font=\bfseries\Large] at (1.95,-1.0) {\color{dimSSE}SDD};
\end{tikzpicture}%
}
\caption{The three necessary and complementary dimensions of SDD. Each is grounded in a different strand of the current research and policy landscape and SDD emerges only where all three overlap.}
\label{fig:sdd-dimensions}
\end{wrapfigure}

SDD is treated in this paper as spanning \textbf{three necessary and complementary dimensions} represented in \autoref{fig:sdd-dimensions}, each grounded in a different strand of the current research and policy landscape.

The \textbf{software and systems engineering (SSE)} dimension is concerned with how defence platforms are designed, procured, and certified. This is the dimension emphasised in the BMVg position paper, which frames SDD primarily as a matter of architecture principles, procurement reform, and the economics of software investment across long platform lifespans, while touching on AI and connectivity requirements only at the level of general capability goals rather than as engineering problems in their own right~\cite{bmvg2023sdd}.

The second dimension is \textbf{AI engineering (AIE)}, concerned with the trustworthiness of the AI-driven and autonomously acting components those platforms increasingly rely on. This reflects a broader shift in defence AI policy towards trustworthiness, transparency, and human oversight as first-order engineering requirements rather than secondary concerns, patent in the NATO revised AI strategy (2024)\footnote{https://www.nato.int/en/about-us/official-texts-and-resources/official-texts/2024/07/10/summary-of-natos-revised-artificial-intelligence-ai-strategy} and echoed in fortiss's own engineering perspective on trustworthy autonomous and cognitive systems~\cite{fortiss2025_trustworthy_ai}. The premise is that the shift towards software-defined platforms is only viable if these components are certifiably safe under adversarial and contested conditions. This framing focuses on mission-critical accuracy, adversarial resilience, uncertainty estimation, and human oversight mechanisms.

The third dimension is a \textbf{connectivity and infrastructure engineering} dimension, concerned with whether the underlying network and computational substrate can deliver information between sensors, AI components, and operators within operationally meaningful bounds~\cite{fortiss2025_6g}. This substrate is best understood through the lens Europe is already converging on for civilian infrastructure: \textit{Connected Collaborative Computing (3C)} networks, integrating telco, edge, and cloud resources into a single federated substrate, as pursued by large-scale European initiatives such as EURO-3C\footnote{https://euro-3c.eu/}. In the SDD context, the same substrate concerns apply, namely routing, data-compute-network orchestration, and in-network processing, rather than the on-device AI inference compute discussed under low-power edge intelligence, which remains part of the AI engineering dimension.

\autoref{tab:crosslayer} summarises the resulting shift from legacy to software-defined paradigms at each layer of the defence system stack.


\begin{table}[htp!]
\centering
\caption{From legacy to SDD defence systems, mapped to the three fortiss vision SDD dimensions.}
\label{tab:crosslayer}
\tiny
\renewcommand{\arraystretch}{1.3}
\begin{tabularx}{\textwidth}{@{}p{2.3cm}XXp{3.3cm}@{}}
\toprule
\textcolor{fortissblue}{\textbf{Layer}} & \textcolor{fortissblue}{\textbf{Legacy}} & \textcolor{fortissblue}{\textbf{SDD}} & \textcolor{fortissblue}{\textbf{Primary Dimension}} \\
\midrule
Platform
  & Hardware-specific, proprietary interfaces
  & Standardised hardware, open APIs, MOSA
  & \dimtag{dimSSE}{Software \& systems engineering} \\
Communication
  & Fixed waveforms and dedicated hardware radios, with protocol stacks coupled to physical hardware
  & SDN, softwarised protocol stacks, virtualised network functions (VNF), programmable data planes
  & \dimtag{dimCON}{Connectivity \& infrastructure engineering} \\
Computation
  & Fixed-function embedded processors
  & General-purpose, high-performance compute, virtualisation, AI accelerators, edge inference
  & \dimtag{dimAI}{AI engineering}\newline\dimtag{dimCON}{Connectivity \& infrastructure engineering (substrate)} \\
Mission logic
  & Hardwired at manufacture
  & Software modules, updateable in-field or over-the-air
  & \dimtag{dimSSE}{Software \& systems engineering} \\
Security
  & Point-in-time validation at delivery
  & Continuous updates, SBOM-tracked supply chain, zero-trust posture
  & \dimtag{dimCROSS}{Cross-cutting compliance (SSE-anchored)} \\
Interoperability
  & Bilateral integration agreements
  & NATO FMN-compliant architectures, open standards
  & \dimtag{dimCON}{Connectivity \& infrastructure engineering} \\
Orchestration
  & Fixed topology, centralised management
  & Software-defined, decentralised, semantics-aware orchestration
  & \dimtag{dimCON}{Connectivity \& infrastructure engineering} \\
\bottomrule
\end{tabularx}
\end{table}

These three dimensions do not operate independently: four convergence drivers, spanning artificial intelligence, autonomy, connectivity, and software and systems engineering reform, are now pulling all three together at once, turning SDD from an engineering possibility into an operational necessity. \autoref{sec:convergence} examines each in turn.
 
\subsection{Convergence Drivers across the Three Dimensions}
\label{sec:convergence}

To make SDD a reality, the three dimensions described need to be worked together. Four convergence factors are driving the need for this dimensional intertwining.

\paragraph{The learning shift: pulling AI trust and certification together.}
AI adoption in defence, from ISR data fusion to real-time decision support and autonomous navigation, is what makes the other two dimensions urgent rather than optional. A model that fuses sensor data across domains needs the connectivity dimension's deterministic, high-bandwidth links to deliver its output in time; a model whose behaviour is learned rather than specified needs the software and systems engineering dimension's certification processes to be reworked, since learned functions replace rule-specified behaviour and existing safety standards do not reason about them~\cite{fortiss_ml_safetycase}. European funding already reflects this shift: the \textit{European Defence Fund (EDF)} 2025 Work Programme launched a dedicated AI technology challenge, and the 2026 Work Programme adds a Digital Transformation call category for AI-based tactical situational awareness using swarms of small robots and drones~\cite{ec2025edf2026}.

\paragraph{Autonomy: the three-dimensional intersection.}
Unmanned systems across ground, air, maritime, and space domains
are expanding the operational envelope of armed forces. A key engineering requirement in this context is that autonomous systems must be both reliably controlled under normal conditions, with human operators retaining decision-making authority and accountability for the use of lethal force, and capable of independent mission-aborting decisions when communications are degraded or denied. This requirement did not arise at scale in earlier generations of CPS engineering\footnote{https://safe-intelligence.fraunhofer.de/en/articles/multi-agent-reinforcement-learning-when-intelligent-systems-must-cooperate-autonomously}~\cite{geisberger2012agendaCPS}, and it cuts across all three SDD dimensions at once: it is an AI-trustworthiness question, a connectivity and infrastructure question, and, through human-machine teaming governance, a software and systems engineering question.

\paragraph{The real-time constraint: binding design to delivery.}
Multi-Domain Operations require data sharing across land, air, maritime, space, and cyber domains simultaneously~\cite{bmvg2023sdd}. This places first-class requirements on the network layers (PHY; MAC; IP): software-controlled spectrum management, deterministic wireless links, and decentralised network
orchestration are necessary properties of the underlying infrastructures. Projects such as SeRANIS\footnote{https://seranis.de/} are developing integrated B5G/6G laboratory environments for this purpose\footnote{https://dtecbw.de/home/forschung/unibw-m/projekt-seranis}. However, the networking layer including deterministic scheduling, semantic compression, and resilient control planes remains the least developed dimension of the current evolution of SDD engineering~\cite{fortiss2025_6g}.

\paragraph{The acquisition bottleneck: certification racing to keep pace.} None of the three described factors delivers an operational advantage if the institutions that acquire, certify, and deploy defence software cannot move at a comparable pace. Existing funding instruments do not, by themselves, supply the engineering process. Procurement models built around a single fixed-scope delivery milestone are structurally mismatched to software expected to change weekly and AI models expected to change daily. Closing that mismatch is a research problem and an administrative one. It requires traceable, modular architectures, certification approaches that evaluate a change rather than re-evaluate an entire platform, and continuous cybersecurity compliance tooling that keeps pace with evolving regulation.

The four convergence drivers share one consequence: the three dimensions can no longer be engineered apart and so the intertwining cannot wait. Each convergence factor exposes the same underlying problem: defence platforms must absorb requirements that evolve continuously, while the hardware carrying them does not. \autoref{sec:lifecycle} names this mismatch the lifecycle paradox.

\subsection{The Lifecycle Paradox}
\label{sec:lifecycle}
\begin{figure}[htbp!]
  \centering
  \includegraphics[width=0.8\textwidth]{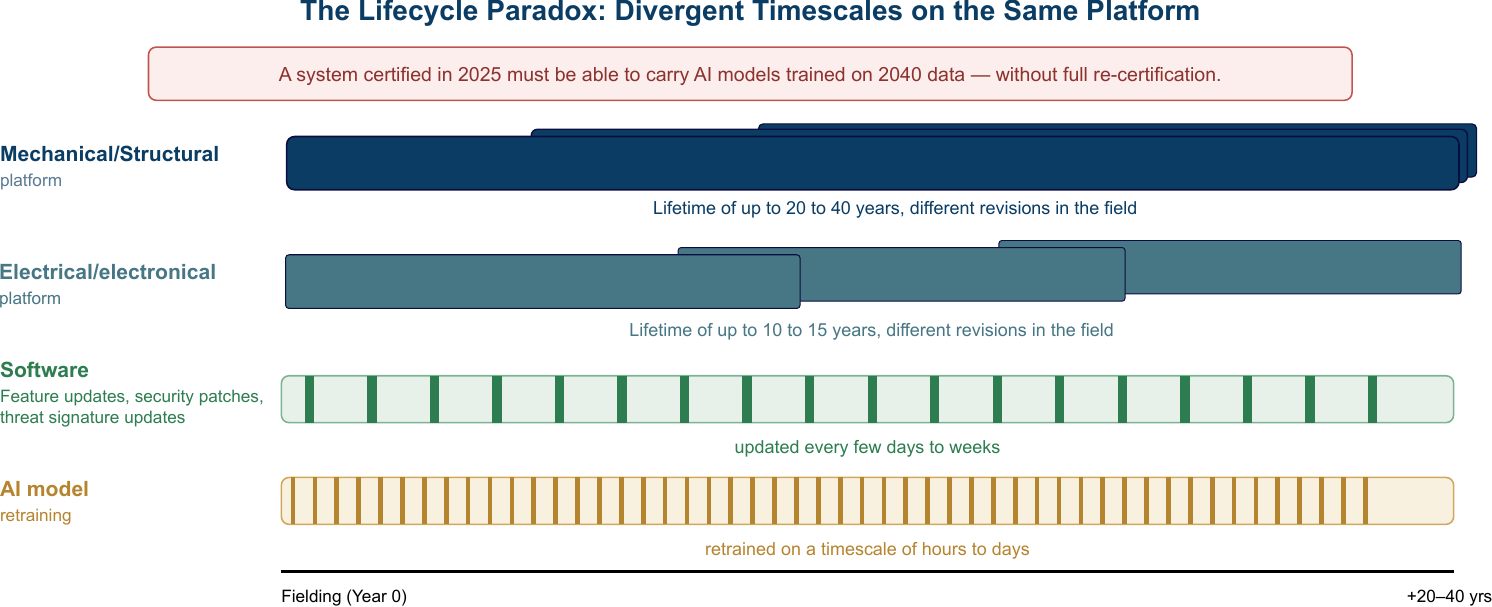}
  \caption{SDD lifecycle paradox: hardware, security, and AI models timing gaps.}
  \label{fig:lifecycle-paradox}
\end{figure}

The structural SDD tension comes from mismatched evolution rates, shown in \autoref{fig:lifecycle-paradox}. Platform lifecycles vary widely. Large, capital-intensive systems are often built around mechanical and structural baselines meant to last decades. Other systems are developed, fielded, and replaced in much shorter cycles. Hardware, compute, sensing, and communication subsystems sit in between still on constrained, multi-year cycles. Software moves the fastest of all. It evolves on timescales of days to weeks, driven by security patches, threat signature updates, hardening measures, bug fixes, and regular feature and capability updates, AI models included~\cite{bmvg2023sdd}.

The foundational literature on CPS engineering has already identified this tension for software-intensive embedded systems~\cite{broy2010cps_acatech, broy2012cps_position}. In defence, this tension increases. Updates must happen under operational conditions and against active adversaries. Safety- and security-critical systems must retain valid qualification and assurance evidence throughout each successive update.

For instance, a system certified in 2025 will need to carry AI models trained on 2040 data without a full re-certification. Security vulnerabilities found after delivery must be fixed in the field, not by returning platforms to depot-level maintenance. Functional updates, a new sensor fusion algorithm or an updated target-identification model, must deploy with documented, bounded impact on the existing assurance case. Updates to electronic subsystems must be assessed for their impact on software execution, interfaces, cybersecurity, and qualification evidence, and the fielded variability in which the results must be handled by software baselines that stay compatible and ensured across subsystem versions. The software supply chain, open-source components, third-party AI frameworks, and update distribution infrastructure alike, becomes a critical attack surface that needs continuous monitoring~\cite{bmvg2023sdd, euCRA2024}, a risk that grows as rapid AI innovation deepens dependency on fast-moving open-source ecosystems.

\subsection{Beyond the Battlefield: Civil and Dual-Use Applications}

The SDD paradigm has applications beyond military systems and operations, reinforcing this paper's civil-to-defence thesis in the opposite direction: capabilities matured for SDD feed back into civilian resilience. The acatech CPS agenda identified disaster response, emergency management, and critical infrastructure protection as primary domains of CPS application~\cite{broy2012cps_position, geisberger2012agendaCPS}, a positioning shared across many other European countries. In the context of SDD, these domains draw on much of the same technology base and face similar engineering challenges.

Two examples illustrate this transfer concretely. In disaster response and crisis management, technologies developed for SDD, including \textit{Unmanned Autonomous Systems (UAS)} for area coverage, AI-fused sensor networks for situational awareness and resilient communications for degraded infrastructure scenarios, are directly applicable to civil emergency response~\cite{fortiss2025_6g,harttunen2025uav}: Germany's disaster control and civil protection frameworks and the EU \textit{Emergency Response Coordination Centre (ERCC)}\footnote{\url{https://civil-protection-humanitarian-aid.ec.europa.eu/what/civil-protection/emergency-response-coordination-centre_en}} are increasingly designed to draw on the same C2 software capabilities used in military contexts. The same holds for the protection of critical infrastructures, where the attack surfaces of energy grids, water systems, and transport control networks increasingly resemble those of military C2 systems~\cite{ec2025roadmap}. Software-defined approaches to network segmentation, anomaly detection, and automated incident response apply in both settings, as recent programmable-network architectures for operational-technology resilience demonstrate~\cite{holik2026ebpf,geisberger2012agendaCPS}.

This civil-military overlap confirms the paper's central thesis: civil-proven technology is not automatically defence-ready. \autoref{challenges} turns to the concrete engineering challenges in all three SDD dimensions, that stand between this shared technology base and genuine deployment under contested, adversarial, or defence-certified conditions.
\section{Engineering for Continuous Change: The Core Challenges}
\label{challenges}  
The lifecycle paradox is not evenly distributed: it lands differently on each of the three SDD dimensions, and each must close a distinct engineering gap before the continuous loop this paper proposes can operate safely. \autoref{sec:challenges-sse} turns to the SSE dimension, where the gap is procedural: acquisition, certification, and supply-chain assurance processes built for a single delivery milestone must instead become iterative and capable of continuously generating evidence. \autoref{sec:challenges-ai} turns to the AIE dimension, where the gap is epistemic: existing safety certification was not built to reason about components whose behaviour is learned rather than specified, or to hold once an adversary is actively present. \autoref{sec:challenges-connectivity} turns to the connectivity and infrastructure engineering (CIE) dimension, where the gap is physical: no certified, trustworthy component delivers operational value if the network carrying its output cannot meet the platform's decision cycle.

\subsection{SSE: From Sequential Acquisition to Continuous Change}
\label{sec:challenges-sse}
Classical defence engineering runs on a straight line: define requirements, design, build, test, certify, field, then touch nothing. That line reflects a hardware-centric world~\cite{broy2010cps_chapter, broy2013engineering_cps}. SDD does not eliminate these activities, but requires them to become more iterative, evidence-driven, and change-aware. The SDD challenges described next are relevant to a rapid, explainable, and adaptive engineering response.

\paragraph{Modular Open Systems Architecture (MOSA).}
MOSA requires standardised, documented interfaces between system modules. This enables third-party software integration, independent module validation, and capability insertion and reuse across platform variants without access to the full platform implementation~\cite{bmvg2023sdd}. MOSA is a prerequisite for DevSecOps in defence. Its realisation at the network layer requires SDN abstractions that make platform APIs accessible across heterogeneous communication environments~\cite{kathiravelu2017sdcps}, a direct dependency on the connectivity and infrastructure engineering dimension. Civilian telecom infrastructure is already moving in this direction: projects such as Euro-3C, Sylva\footnote{https://sylvaproject.org/}, CODECO\footnote{https://he-codeco.eu/} are backed by major European carriers and vendors and are  building open, standardised telco cloud stacks precisely to eliminate the interface fragmentation MOSA targets in defence, though for orchestration and compute rather than the tactical waveform layer. At the engineering level, MOSA also requires explicit architecture models that capture module boundaries, interface assumptions, and dependencies, and that provide traceability links to underlying requirements and assurance evidence.

\paragraph{DevSecOps for defence.}
Continuous integration, delivery and security assurance pipelines, adapted from commercial software practice for defence assurance requirements, are required to reduce the time from software change to fielded capability from years to weeks, while maintaining verifiable dependability properties at each iteration~\cite{bmvg2023sdd}. For SDD, these pipelines must continuously generate and maintain assurance evidence, as an integral part of normal engineering workflows, rather than as a separate, retrospective activity performed once development stops.

\paragraph{Incremental (continuous) certification.}
Traditional certification treats qualification as a labour-intensive, lengthy audit activity built around normative reference models. That approach does not cope with fast and adaptive engineering environments, where software changes continuously. Incremental certification, where a change to one module triggers re-evaluation of only the affected assurance arguments, requires new modular approaches in line with the system decomposition MOSA already supports. This in turn requires traceability-supported change-impact analysis, automated retrieval of assurance evidence for software updates and architecture changes, support for evolving regulatory frameworks, and human-in-the-loop mechanisms, not least for reasons of accountability and liability.

\paragraph{Software supply chain security and SBOM.}
Modern defence software stacks incorporate large numbers of open-source components and libraries, commercial AI frameworks, and third-party middleware. The dependency graph of a representative defence software stack can contain thousands of components from multiple vendors~\cite{bmvg2023sdd}. CPS of this scale and complexity were already identified in the agendaCPS study as requiring new approaches to dependency management and trustworthy composition~\cite{geisberger2012agendaCPS}. SBOM obligations under the EU CRA~\cite{euCRA2024} provide a regulatory baseline; their defence-specific extension — classified dependency tracking, continuous provenance verification — remains an open engineering problem.

\subsection{AIE: Trust Under Uncertainty and Contact with Adversaries}
\label{sec:challenges-ai}
Two questions define the core challenges of this dimension. First: how does a component behave once it meets an adversary actively trying to break it, rather than the benign conditions under which it was tested? Second: once a system acts autonomously, who stays responsible for what it does? Neither question has a settled answer in classical safety engineering, and both demand more from AI components than from traditional, deterministic software.

\paragraph{Trustworthy AI.}
Existing safety certification frameworks are designed for deterministic systems; they do not accommodate the non-deterministic failure modes of AI components. Defence deployments expose AI systems to adversarial scenarios~\cite{jedrzejewski2024adversarial} and operational conditions that can deviate from field tests and undermine assurance arguments built on testing alone. Explainability requirements add a further demand: defence-critical decisions must support human review. Trustworthiness in these components therefore carries a higher bar than in more traditional systems. Future autonomous systems will also require mechanisms for continual adaptation to evolving operational environments, creating assurance challenges beyond those posed by static AI models.

\paragraph{Graceful degradation and operational continuity.}
Classical safety engineering emphasises fail-stop behaviour: the system halts when it encounters conditions outside its validated envelope. This paradigm suits many civilian CPS applications~\cite{broy2012cps_position}. In defence and crisis-management contexts, continued operation in a degraded state is often preferable to halting because inaction can itself be a mission failure. This requires runtime adjustment of risk-acceptance thresholds and the ability to transition between pre-mission, in-mission, and post-mission operational modes as first-class architectural concerns. As in the civilian context, fail-operational behaviour may be required for certain safety-critical system classes — manned airborne systems in particular, where loss of function could lead to catastrophic consequences. Realising these transitions as first-class architectural concerns also depends on the traceable modular architectures named under software and systems engineering.

\paragraph{Human-machine teaming and autonomy governance.}
\textit{International Humanitarian Law (IHL)} requires that meaningful human control be maintained over the use of force. This is an engineering constraint and a legal one: the system must provide mechanisms for operators to intervene, override, or abort at operationally relevant timescales. These mechanisms must be reflected in the system architecture, user interfaces, documentation, runtime monitoring, and assurance case.

\subsection{CIE: When the Network Is the Bottleneck}
\label{sec:challenges-connectivity}

A certified AI component that cannot deliver its output to a C2 node within the required decision cycle provides no operational value, no matter how trustworthy that component is in isolation. Network conditions are therefore not a deployment detail but an architectural and assurance-relevant concern, affecting function allocation, fallback behaviour, operational modes, runtime reconfiguration, and the validity of safety and security arguments built elsewhere. Three challenges carry a particular weight for SDD.

\paragraph{Deterministic delivery at the tactical edge.}
Multi-Domain Operations require sensor data, C2 commands, and AI-derived outputs to traverse heterogeneous radio environments with bounded, predictable latency~\cite{bmvg2023sdd}. Extending \textit{Time Sensitive Networking (TSN)} determinism from wired backbones to mobile multipoint wireless environments remains an open engineering problem, with time-aware scheduling and tight synchronisation across multi-AP Wi-Fi~6/7 as the key requirements~\cite{fortiss2025_6g,sofia2024detnetwifi}.

\paragraph{Bandwidth-constrained, meaning-preserving transmission.}
Tactical radio links operate under severe and variable bandwidth limitations. Semantic communications, i.e., transmitting compressed representations that preserve intent rather than raw data streams reduces what must be sent to the minimum the receiver needs to reconstruct operationally relevant information~\cite{fortiss2025_6g}, directly realising the CPS property of adaptive, resource-sensitive communication~\cite{geisberger2012agendaCPS}.

\paragraph{Resilience when infrastructure is degraded, denied, or absent.}
Centralised network management is a single point of failure in a jamming, cyberattack, or physical destruction. Cognitive, decentralised orchestration — where network services adapt and re-route autonomously without central coordination — addresses continuity under active interference~\cite{fortiss2025_6g}. Denied, Disrupted, Intermittent, and Limited (DDIL) conditions push this even further: links may be severed for seconds to hours by jamming, orbital geometry, or terrain masking, and connection-oriented protocols fail silently~\cite{dtn_rfc9171}. \textit{Delay/Disruption-Tolerant Networking (DTN)} store-carry-forward semantics decouple delivery from contemporaneous end-to-end connectivity; publish-subscribe decouples producers from consumers, so state updates can be retrieved opportunistically across whatever path exists. Both align with data-oriented in-network caching approaches such as \textit{Named Data Networking (NDN)}~\cite{zhang2014ndn} where any holding  node serves a cached update regardless of producer reachability.

\section{A Continuous Engineering Loop for SDD}
\label{sec:capabilities}

\begin{figure}[htp!]
  \centering
  \includegraphics[width=1.0\textwidth]{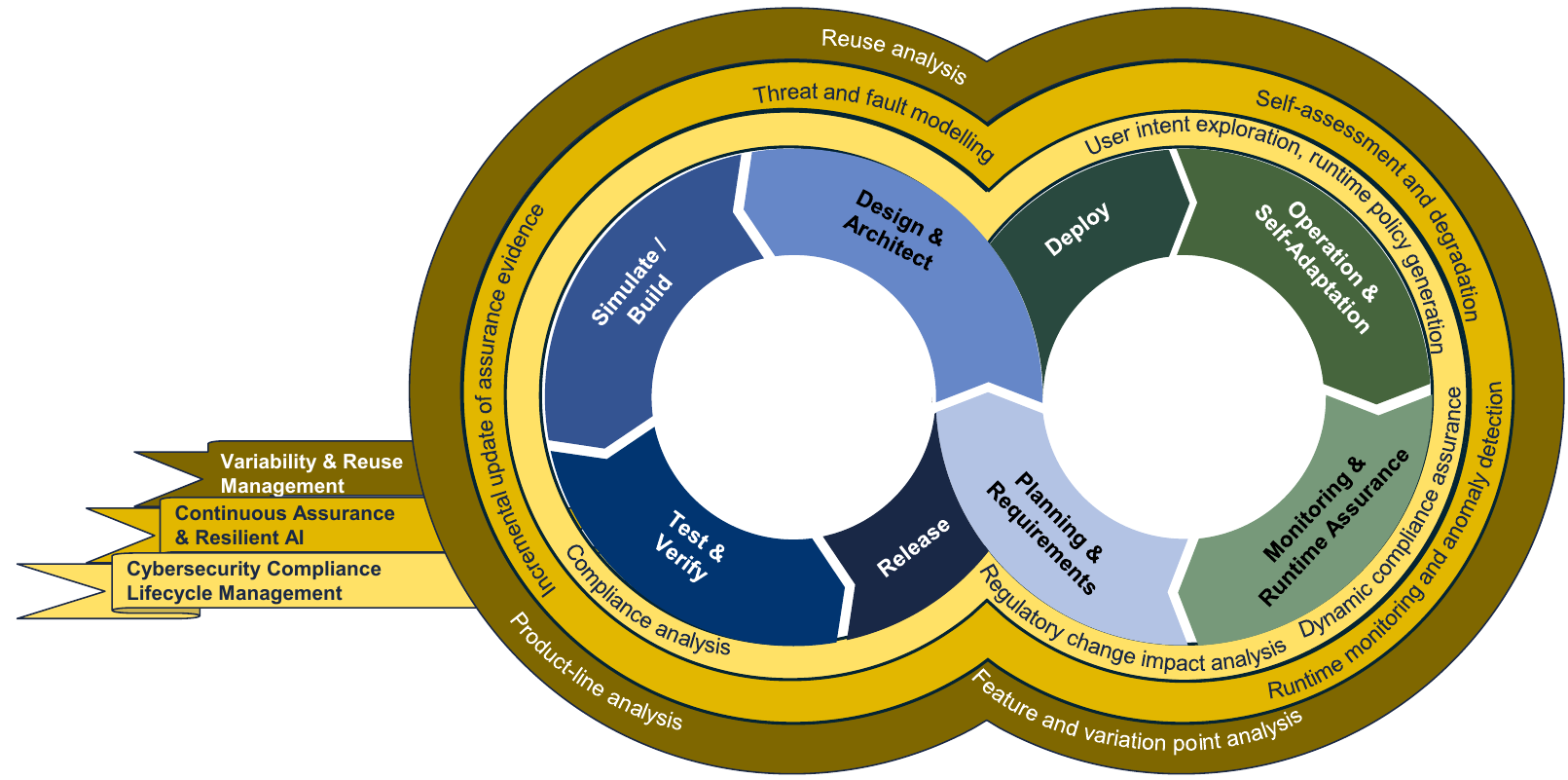}
  \caption{Capability areas for SDD as stages and cross-cutting concerns of a continuous engineering loop.}
  \label{fig:capabilities-pipeline}
\end{figure}

The challenges in \autoref{challenges} intertwine: an AI model built without compliance tracking cannot be certified quickly; a component tested only under nominal conditions cannot be trusted under adversarial conditions; and a trustworthy AI component that cannot reach its operator in time provides no operational value, regardless of how accurate it is. The capabilities discussed in this section overcome these challenges through a continuous, DevOps-inspired engineering loop represented in \autoref{fig:capabilities-pipeline}. 

Four core capabilities are positioned along the recurring activities of the loop: \textbf{Model-Based Systems Engineering for Design and V \& V Front-Loading} (\autoref{sec:mbse}), the backbone of the digital engineering process; \textbf{simulation-based testing} (\autoref{sec:testing}), for early verification and validation; \textbf{tactical and resilient connectivity} (\autoref{sec:infrastructure}), for deployment and resilient operation under contested conditions; and \textbf{low-power edge execution} (\autoref{sec:nc}), for operation at the tactical edge. 

Three additional capabilities are cross-cutting concerns that span the loop continuously: \textbf{cybersecurity compliance lifecycle management} (\autoref{sec:compliance}), \textbf{continuous assurance and resilient AI} (\autoref{sec:continuous-assurance-resilient-ai}), and \textbf{variability and reuse management} (\autoref{sec:variability}). 

During the development (\textit{Dev}) phase of the loop, these capabilities enable agile, iterative design while staying compatible with the V-shaped development and assurance processes that standards require. This reduces risk on complex projects and supports efficient responses to requirements changes as they arise in multi-stakeholder work. 
A traceability-supported, artefact-based engineering approach links requirements, architectural views, interfaces, implementation artefacts, simulation models, test artefacts, analysis results, and assurance evidence. This enables front-loading of architectural trade-offs, including the selection and dimensioning of dependability mechanisms such as containment, recovery, and graceful degradation, and supports simulation- and test-based V\&V of their dynamic behaviour. When requirements, architectures, code, analysis results, or assumptions change, the same traceability helps identify affected evidence and assurance claims. Fault, threat, context, architecture, and interface models can then be used to derive test scenarios that cover edge cases, varying operating conditions, worst-case situations, and adversarial conditions beyond nominal behaviour.

During the Operational (\textit{Ops}) phase, tactical and resilient connectivity carries updates and operational data under degraded or contested conditions, while low-power edge execution keeps software and AI functions within the resource, latency, and energy limits of embedded platforms. Runtime monitoring and self-assessment flag when the system must shift into a degraded mode; dynamic compliance assurance keeps behaviour aligned with regulatory and mission constraints. Observations, incidents, and new evidence then feed back into Dev, triggering requirement updates, re-testing, or revised assurance arguments as needed, closing the loop between the two phases.

\subsection{Core Capabilities}
\subsubsection{Model-Based Systems Engineering for Design and V \& V Front-Loading}
\label{sec:mbse}
SDD systems cannot be engineered and maintained efficiently through textual requirements and implementation artefacts alone. Moving from requirements to design and V\&V requires a methodological approach that makes explicit what the system shall achieve, where and under which conditions it operates, and under which constraints it must be designed, analysed, validated, and updated.

\textit{Model-based systems engineering (MBSE}) provides that backbone. Independently of the modelling language or tool, MBSE offers an artefact-based approach in which models from different viewpoints are gradually refined as the system is decomposed into subsystems, components, and interfaces, with assumptions, constraints, and design objectives made explicit. Models capturing usage contexts, operators, and external actors define the environment the system must operate in; functional and non-functional requirements models capture the problem space; the design space defines possible realisations, shaped by applicable regulations, standards, and platform constraints. Traceability links between these versioned artefacts establish a digital thread in which engineering decisions are documented and associated with supporting evidence such as simulation results, test results, analysis results.
Models enable system-level analyses in early engineering phases, which matters particularly for SDD, where architectural decisions must account for modular capability insertion, software and AI upgradability, and constrained, contested connectivity all at once. Hazard, risk, fault, threat, and loss analyses, e.g., \textit{Hazard Analysis and Risk Assessment (HARA)}\footnote{A systematic method for identifying hazards and assessing associated risk levels in a system, standardised for road vehicles in ISO~26262~\cite{iso26262} and widely adapted across other safety-critical domains.}, \textit{System-Theoretic Process Analysis (STPA)}\footnote{A hazard analysis technique based on systems theory rather than component failure, treating accidents as arising from inadequate control rather than only from component faults~\cite{stpa_handbook}.}, \textit{Fault Tree Analysis (FTA)}\footnote{A deductive, top-down method that models how combinations of lower-level faults or events can lead to a specified undesired system-level failure, standardised in IEC~61025~\cite{iec61025}.}, \textit{Threat Analysis and Risk Assessment (TARA)}\footnote{A structured method for identifying cybersecurity threats to a system and assessing the associated risk, standardised for road vehicles in ISO/SAE~21434~\cite{isosae21434} and adapted for other connected and autonomous systems.}, and \textit{Failure Mode and Effects Analysis (FMEA)}\footnote{A bottom-up, inductive method that systematically evaluates potential failure modes of system components and their effects on system behaviour, standardised in IEC~60812~\cite{iec60812}.} identify how faults, adverse inputs, or component failures affect system behaviour, and feed the design of dependability mechanisms such as containment, recovery, and graceful degradation.

Semantically rich, traceable models also let formal methods, optimisation, and \textit{generative AI (genAI)} explore design alternatives early: e.g., evaluating alternative allocations of autonomy functions between edge nodes and C2 infrastructure, assessing fallback behaviour under degraded communication, or checking whether an update affects timing, resource, or dependability assumptions. As \autoref{sec:testing} discusses, these same models front-load (V\&V) through simulation-based testing, especially for emergent properties visible only in the integrated system, e.g., control stability, timing behaviour, robustness of dependability mechanisms.

\subsubsection{Simulate \& Test: Simulation-based Testing of Autonomous Systems}
\label{sec:testing}
SDD changes what testing is for: verification and validation become a continuous capability accompanying software throughout its operational lifetime. Every update, retraining run, reconfiguration, or sensor replacement can change system behaviour, and each needs fresh evidence that the system remains safe, secure, and effective.

\textbf{System-level simulation-based testing} provides that evidence rapidly and safely before updates reach operational platforms. Integrating executable system models with environment simulators enables verification of emergent properties invisible at component level. Certification evidence is built through progressively higher-fidelity testing, i.e.,  Software-in-the-Loop, Hardware-in-the-Loop, System-in-the-Loop, and ultimately real-world trials, requiring coordinated multi-fidelity pipelines that keep lower-fidelity evidence predictive of operational behaviour.

Manual construction of representative defence scenarios is infeasible given the scale of the operational design space (e.g., environmental conditions, hardware failures, adversarial behaviour, mission objectives). \textbf{Search-based testing} explores this space automatically, generating critical scenarios derived from parametrised fault, threat, context, and interface models to expose weaknesses and previously unknown failure modes.

A further challenge is keeping simulated evidence predictive of reality: differences in sensor characteristics, environment, and hardware execution create a simulation-to-reality (\textbf{Sim2Real}) gap. Continuous verification requires systematic comparison across testing modalities and mechanisms to determine when simulated failures are likely to transfer to operational deployments. Foundation models can support scenario generation, test oracle construction, fault localisation, and regression testing combined with search-based optimisation, letting engineers analyse increasingly complex autonomous systems while maintaining the evidence continuous assurance and incremental certification require.

\subsubsection{Tactical and Resilient Connectivity}
\label{sec:infrastructure}

Tactical networks combine requirements that civilian networks rarely face together: contested spectrum and active jamming, heterogeneous mobile CPS sets, hard QoS for mixed traffic classes, and resilience against destruction or cyberattack of management infrastructure. Three regimes structure the challenge: \textbf{deterministic} operation when infrastructure is present, \textbf{opportunistic} operation when it is absent or partitioned, and \textbf{delay-tolerant} operation under severe QoS constraints. A fourth concern, \textbf{data distribution middleware}, cuts across all three: it determines how data reaches consumers with appropriate QoS in a decoupled, infrastructure-adaptive way.

\paragraph{Deterministic operation.} A network with bounded nominal-condition latency that degrades silently under jamming or node loss fails the operational requirement: safety-critical traffic must arrive within its deadline, as after that it cannot be considered. IEEE \textit{Time Sensitive Networking (TSN)} standards and IETF \textit{Deterministic Networking (DetNet)} standards provide the architectural basis for deterministic delivery where infrastructure is present, with recent controller-plane standardisation extending that basis across heterogeneous domains~\cite{rfc9938}. The open problem is the \textit{Denied, Disrupted, Intermittent, and Limited (DDIL)} boundary: resynchronisation across the different devices in case connectivity gets lost to reference clocks, and a handoff to DTN store-carry-forward~\cite{dtn_rfc9171} when TSN guarantees cannot be upheld, so that deadline semantics survive the opportunistic transport layer. This handoff, and its inverse on link restoration, remain unsolved; \autoref{sec:contrib-infrastructure} presents fortiss evidence toward closing this gap.

\paragraph{ICN over DTN for DDIL.} DTN (RFC~9171~\cite{dtn_rfc9171}) addresses jamming, orbital geometry, and terrain masking through store-carry-forward at the Bundle Protocol layer; DTNMA (RFC~9675~\cite{rfc9675}) formalises autonomous local control of DTN nodes for forward-deployed units without C2 connectivity. NDN~\cite{zhang2014ndn}, as a reference open architecture for \textit{Information centric Networking (ICN)}forwards \textit{Interests} toward any satisfying source and caches responses in-network. Hence, this data-centric networking paradigm is well suited to military mobility and link intermittence, with demonstrated gains over IP in WIN-T/ADNS emulation~\cite{etefia2012milcom}. A recurring approach for space and tactical communications combines both: NDN forwarding when links are available, DTN-bundle encapsulation under DDIL, bridging to PubSub middleware (DDS\footnote{\url{https://www.omg.org/omg-dds-portal/}}, MQTT~\cite{mqtt_nato2018}) and, where bandwidth binds, semantic encoding. Despite these advances, open questions still remain in the context TSN-to-DTN handoff standardisation, NDN-DDS convergence, securing ICN-DTN relay chains against node compromise and Interest poisoning, and agentic contact-schedule management across space and ground segments~\cite{dtn_rfc9171}. \autoref{sec:contrib-infrastructure} presents fortiss evidence for this combined approach.

\subsubsection{Low-power Edge Execution}
\label{sec:nc}
As operational environments grow more dynamic and contested, defence systems must do more than execute pre-trained models. They must continuously adapt to new environments, evolving threats, and unforeseen conditions, often with limited bandwidth, restricted or denied connectivity, and no guaranteed access to centralised compute. This is an AI engineering challenge as much as a hardware one: SDD requires not only intelligence at the edge but learning at the edge: perceiving, inferring, and adapting directly on embedded platforms, under strict size, weight, and power constraints.

Neuromorphic computing offers the most promising path toward this. The DARPA's SyNAPSE programme\footnote{\url{https://www.darpa.mil/research/programs/systems-of-neuromorphic-adaptive-plastic-scalable-electronics}} identified brain-inspired computing as a path toward systems that learn from experience within stringent energy budgets, still highly relevant for next-generation SDD, and consistent with NATO's own emphasis on AI-enabled autonomy, human-machine teaming and operational resilience~\cite{nato2024ai}. \autoref{sec:contrib-nc} presents fortiss evidence for this approach.

\subsection{Cross-cutting Capabilities}
Three further capabilities run continuously, underneath every core capability from design through operation.

\subsubsection{Continuous Cybersecurity Compliance Lifecycle Management}
\label{sec:compliance}
Regulatory obligations on software-intensive defence systems are numerous, overlapping, and constantly changing. They address not only product features but AI assurance and export control across multinational programmes. Three capabilities make continuous compliance workable at this scale. Compliance has to start by design, not as an afterthought. Conformance needs verifying both at development time and continuously at runtime. And for systems operating autonomously, runtime intent exploration and policy enforcement keep behaviour within legally and ethically sanctioned boundaries as operational contexts evolve.

\subsubsection{Variability and Reuse Management}
\label{sec:variability}
SDD systems rarely exist as a single, uniform baseline. Subsystem versions, software baselines, AI model versions, and configuration profiles often coexist. Therefore, each new capability increment should not require engineering from scratch. Traceability between requirements, architecture, interfaces, and assurance evidence lets product-line and variation-point analysis determine which combinations remain valid. That reduces effort at design time — requirements, architecture fragments, and test scenarios get reused wherever their assumptions still hold — and constrains what reaches the field, so updates land only where compatibility and assurance conditions are satisfied.

\subsubsection{Continuous Assurance and Resilient AI}
\label{sec:continuous-assurance-resilient-ai}
Testing against nominal and edge-case scenarios does not suffice to make a component resilient. An adversary can probe for exactly the input conditions testing did not cover, e.g., spoofed sensors, decoys, electronic interference, biased training data. Hardening a component means detecting and responding to that manipulation across its whole lifecycle, from threat modelling at design time to anomaly detection at runtime. This closes part of the trustworthy-AI gap previously identified: a certification argument built only on best-case conditions testing will not hold once an adversary is assumed present.

Resilient AI implies that the AI system remains within its validated envelope. This requires continuous self-assessment of compute, energy, infrastructure metrics such as bandwidth, and also confidence. The self-assessment becomes extremely important in situations where there is a gap in human control, e.g., due to lack of connectivity, or due to any other boundary condition. Pre-specified fallback behaviour then needs to take over — bounded, cautious by default, and built to hand control back to the operator the moment connectivity or confidence returns.

\section{From Civilian Research to Defence-Ready Capability}
\label{sec:contributions}
\begin{figure}[htbp!]
  \centering
  \includegraphics[width=\textwidth]{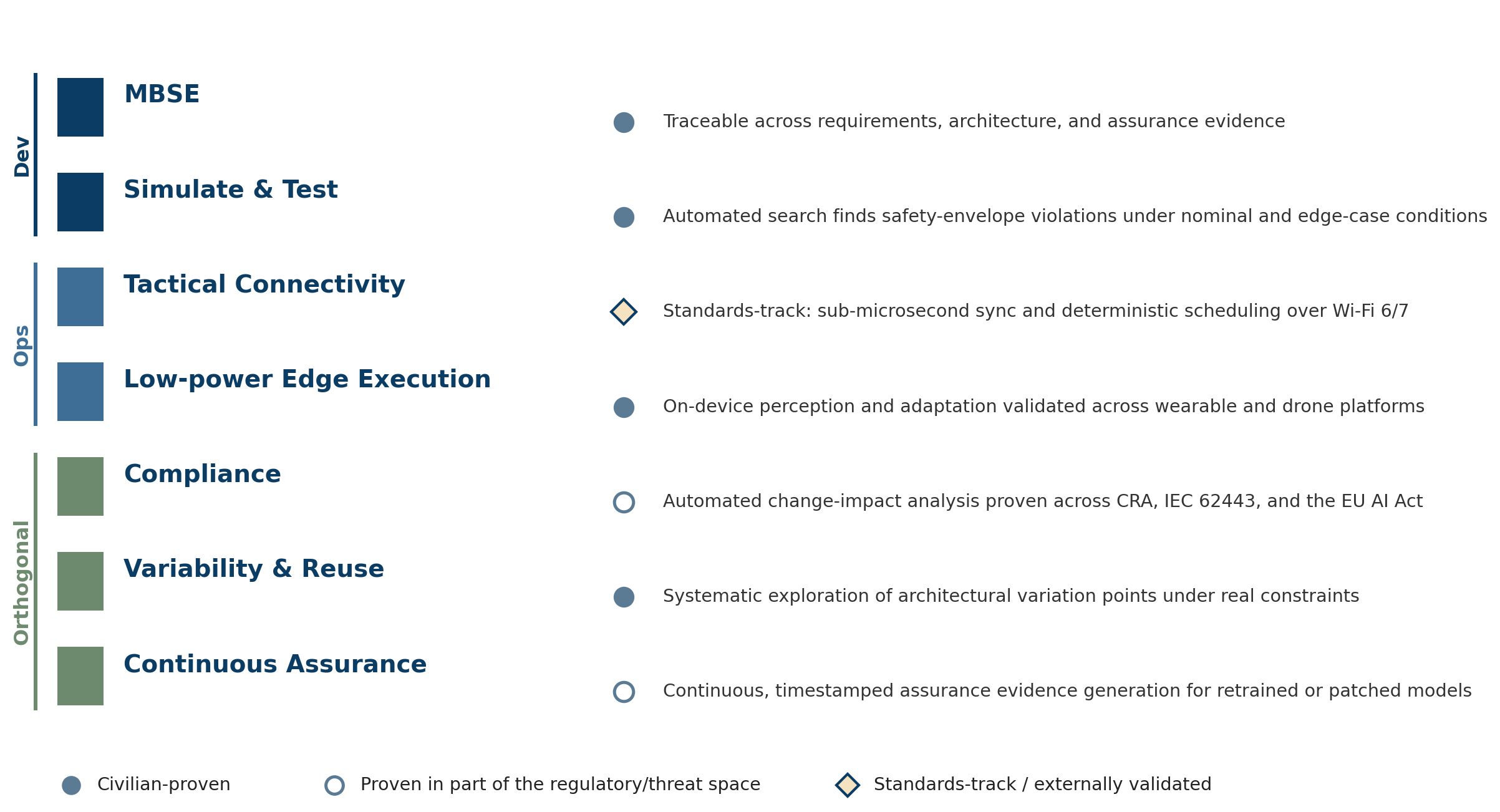}
  \caption{SDD capabilities, and transfer maturity status, civilian to defence.}
  \label{fig:asset-positioning}
\end{figure}

\autoref{sec:capabilities} named seven capabilities a continuous SDD engineering loop requires, largely evidenced so far in civilian and dual-use settings. This section asks, for each of the four core capabilities, the same question: what civilian-domain evidence already exists that a given engineering approach works, and what specifically is missing before it can be trusted under contested, defence conditions? The guiding question throughout is not \textit{"what must be built"} but \textit{"what already exists that gets us there, and where does it still fall short?"}.
Each subsection follows the continuous engineering loop from \autoref{sec:capabilities} and is illustrated with representative fortiss results  chosen because their underlying method is directly transferable, set alongside a comparable effort from outside fortiss to show the direction is an active field, rather than a single-institution claim. \autoref{fig:asset-positioning} summarises the resulting maturity picture, per capability and per \textit{technology readiness level (TRL)}, across all seven capabilities, core and cross-cutting.

\subsection{Model-Based Systems Engineering}
\label{sec:contrib-mbse}

\noindent\textbf{Concept.} \autoref{sec:mbse} calls for an engineering methodology built on semantically rich, traceable models that enables automation of design and V\&V tasks across model viewpoints and refinements.

\noindent\textbf{Why it matters for SDD.} SDD architectures continuously evolve and must satisfy conflicting design constraints, such as available resources, criticality, and timing as well as objectives, such as size, weight, power, and reliability.

\noindent\textbf{Illustrative evidence.} This methodological direction has a well-established civilian track record at fortiss. As one open-source instantiation of that foundation, the fortiss AutoFOCUS3\footnote{\label{fn:af3}\url{https://af3.fortiss.org}} MBSE platform supports SPES~\cite{Bayha2025}-oriented architecture modelling and front-loads design and V\&V through formal analysis and co-simulation; it has been used to validate architecture trade-off analysis and synthesis under resource, timing, and dependability constraints in automotive and aerospace settings~\cite{Eder2020,Dantas2023,Terzimehic2023,Munaro2025}.\footnote{Developed and maintained at fortiss; see \nameref{app:labs} in Annex I, entry Mobility Lab.} Comparable MBSE-for-defence efforts are under way elsewhere, including the NATO Architecture Framework, which provides methodology and viewpoints for military architecture descriptions~\cite{nato2026naf}, and the EU-funded DART project, which combines MBSE, digital twins, and system-of-systems interoperability for military engineering contexts~\cite{edf2025dart}.

\noindent\textbf{Maturity and gap.} TRL~5-6, open-source and validated in civilian CPS domains. The assets support architecture modelling, automated trade-off analysis, and scoped change-impact analysis; the defence gap is to demonstrate them together with continuous assurance (\autoref{sec:contrib-assurance}) on defence-representative architectures and toolchains as demanded by \autoref{sec:challenges-sse}.

\subsection{Simulate \& Test}
\label{sec:contrib-testing}
\noindent\textbf{Concept.} \autoref{sec:testing} calls for testing that runs continuously rather than as a single nominal-condition gate, including systematic search for edge-case and adversarial failure modes.

\noindent\textbf{Why it matters for SDD.} A model validated only under nominal conditions provides no evidence of how it behaves once an adversary is actively present; a certification argument built on that evidence stops holding the moment conditions turn contested.

\noindent\textbf{Illustrative evidence.} Search-based testing automates exactly this kind of scenario discovery. OpenSBT, an open-source framework for search-based testing of automated driving systems, has been validated on automated-emergency-braking functions in both the CARLA simulator and the high-fidelity Prescan environment with industrial partner DENSO~\cite{sorokin2024opensbt}, and extended with fault-injection test-case generation~\cite{munaro2025faultinjection} and simulator-ensemble approaches that narrow the sim-to-real gap~\cite{sorokin2026ensembles}.\footnote{Maintained at fortiss; repository listed in \nameref{app:labs}.} The search mechanism itself is domain-agnostic: redirected from braking-distance objectives to a UAS navigation or target-identification model, with jamming or spoofing as the fitness objective instead of collision distance, the same pipeline could surface degraded-GPS or adversarial-perturbation failure modes in simulation before fielding. A comparable, defence-native effort exists outside this line: DARPA's GARD program develops evaluation methodology for adversarial attacks on ML-based vision and autonomy systems\footnote{https://www.darpa.mil/research/programs/guaranteeing-ai-robustness-against-deception}, evidence that this class of testing gap is an active, recognised problem rather than a hypothetical one.

\noindent\textbf{Maturity and gap.} TRL~5-6, open-source. The search mechanism is directly transferable to defence environments. Redirected from braking-distance objectives to a UAS navigation or target-identification model, the same automated search for input conditions that push a model outside its safety envelope could surface degraded-GPS or adversarial-perturbation failure modes in simulation, before fielding, far more cheaply than discovering them in the field. Validation to date covers nominal-fault and edge-case scenarios. Jamming, spoofing, and sensor denial, the actively adversarial conditions defence autonomy certification requires are next steps.

\subsection{Tactical and Resilient Connectivity}
\label{sec:contrib-infrastructure}

\noindent\textbf{Concept.} \autoref{sec:infrastructure} named three regimes a tactical network must support: deterministic operation while infrastructure is available, opportunistic operation once it is not, and delay-tolerant operation under DDIL.

\noindent\textbf{Why it matters for SDD.} Sensor-to-shooter data that arrives a second late is worthless however accurate the AI behind it; a network that degrades silently rather than gracefully turns a bandwidth problem into a safety problem.

\noindent\textbf{Illustrative evidence.} fortiss is developing the deterministic regime directly: bringing IEEE TSN determinism to Wi-Fi~6/7 through wireless-native mechanisms such as \textit{Fine Time Measurement (FTM)} synchronisation, \textit{Target Wake Time (TWT)}-based scheduling compatible with TSN, and multi-AP interference mitigation~\cite{mohan2022ftm,mohan2023ftm_multiap,schn2022,sofia2024detnetwifi}. This work has also contributed to IETF DetNet/RAW standardisation~\cite{sofia2024ietfraw,rfc9912}, therefore becoming ready for a multi-vendor effort\footnote{Fortiss contributions documented in \nameref{app:labs}, under IIoT Lab.}.

For the delay-tolerant regime, fortiss has demonstrated a store-carry-forward and in-network-caching architecture combining DABBER with NDN-over-DTN for intermittent connectivity in emergency and rescue scenarios~\cite{sofia_dabber,sofia2021ndndtn}. The same store-carry-forward principle is validated well beyond this line of work: NASA/JPL's Interplanetary Overlay Network has operationally delivered Bundle Protocol traffic under equivalent disruption-tolerant conditions for deep-space and tactical links\footnote{https://www.nasa.gov/directorates/somd/space-communications-navigation-program/interplanetary-overlay-network/}.

What remains open is the handoff between these regimes: resynchronising deterministic guarantees after a DDIL interruption, and its inverse on link restoration, is unsolved in both the civilian and defence literature.

\noindent\textbf{Maturity and gap.} TRL~6-7, civilian and standards-track, per the NASA-originated technology readiness scale as adopted for EU-funded research, with validation covering nominal and industrial interference conditions. What is missing is validation of the deterministic-to-opportunistic handoff itself under representative jamming, not of either regime in isolation.

\subsection{Low-power Edge Execution}
\label{sec:contrib-nc}

\noindent\textbf{Concept.} \autoref{sec:nc} argued SDD needs learning at the edge: perceiving, inferring, and adapting directly on embedded platforms under strict size, weight, and power limits.

\noindent\textbf{Why it matters for SDD.} A soldier-worn or platform-mounted system that depends on a live link to remote compute stops working the moment that link is denied, and contested environments deny exactly that link.

\noindent\textbf{Illustrative evidence.} Neuromorphic, event-driven computing is one civilian-validated route to this. Spike-based hardware has demonstrated order-of-magnitude gains in energy efficiency and latency over conventional inference, including on-chip continual learning~\cite{hajizada2026clane}, applied to wearable, real-time action recognition in the EMMANÜELA project.\footnote{Developed at the fortiss Neuromorphics Lab; see \nameref{app:labs}.} Complementary event-based sensing work provides the perception side of the same perceive-infer-adapt loop that \autoref{sec:nc} calls for, and has also been applied to low-latency, event-based drone vision for asset tracking~\cite{neumeier2025eevact,vonarnim2024eventdriven}. This is not an isolated research direction: Intel's Loihi research chips and BrainChip's Akida target the same event-driven, low-power profile for embedded deployment\footnote{https://www.intel.com/content/www/us/en/research/neuromorphic-computing-loihi-2-technology-brief.html}, indicating a maturing hardware ecosystem rather than a single lab's prototype. Further internally-explored applications, including satellite fault detection and gesture-based vehicle control, are at an earlier stage and are not yet independently published, and are therefore not presented here as evidence.

\noindent\textbf{Maturity and gap.} TRL~5-6, civilian-validated, per the NASA-originated technology readiness scale as adopted for EU-funded research. The maturity claim covers inference and on-chip continual learning under benign conditions; robustness of the on-device model itself to adversarial sensor input, the class of gap DARPA's GARD program targets for ML components generally\footnote{https://www.darpa.mil/research/programs/guaranteeing-ai-robustness-against-deception} requires further analysis.

\subsection{Cross-cutting Capabilities}
\subsubsection{Continuous Cybersecurity Compliance}
\label{sec:contrib-compliance}

\noindent\textbf{Concept.} \autoref{sec:compliance} argued compliance has to run continuously across the loop, instead of waiting for a concrete milestone audit.

\noindent\textbf{Why it matters for SDD.} A defence platform answers to overlapping regimes at once, national security and safety standards, NATO interoperability requirements, export control, each evolving on its own schedule. An update that is compliant today can fall out of compliance tomorrow without anyone noticing until the next audit.

\noindent\textbf{Illustrative evidence.} fortiss's compliance research grew out of long-standing collaborations between academia and industry on continuous security engineering~\cite{moyon2024industrial,moyon2025aligning,angermeir2024towards}, already synthesising various and potentially overlapping regulatory norms and standards, performing automated extraction and change-impact analysis of evolving regulatory obligations~\cite{elahidoost2026investigating}, extended to complex deployment environments~\cite{landeck2025} and operational security-finding management~\cite{voggenreiter2024automated,kosenkov2025remapping}. It feeds directly into stress-testing the compliance logic behind an EU AI Act Compliance Checker\footnote{https://artificialintelligenceact.eu/assessment/eu-ai-act-compliance-checker/} and early agentic tooling for CRA- and IEC~62443-related lifecycle management.

\noindent\textbf{Maturity and gap.} TRL~5-6, civilian and dual-use, proven across CRA and immediately related IEC~62443 as well as the EU AI Act, next to other domain-specific norms (e.g. FinTech). The same change-impact approach is already well suited to track a defence platform's obligations across national safety standards, NATO interoperability requirements, EU CRA SBOM obligations, and export control at once, flagging when a proposed update changes the platform's regulatory exposure before it is fielded rather than during a subsequent audit.

\subsubsection{Variability and Reuse Management}
\label{sec:contrib-variability}
\noindent\textbf{Concept.} \autoref{sec:variability} argued SDD systems need systematic control over valid feature, subsystem, and configuration combinations, while reusing requirements, components, and assurance evidence across variants and products.

\noindent\textbf{Why it matters for SDD.} A fielded SDD family will accumulate different subsystem versions, software baselines, AI models, and configurations. Some combinations may violate required properties, be incompatible, or lack valid test and assurance evidence. Re-engineering similar artefacts for every version or product does not scale.

\noindent\textbf{Illustrative evidence.} The product-line analysis~\cite{Bayha2024} available in AutoFOCUS3\footref{fn:af3} determines compatible combination of features and versions~\cite{Bayha2024}. Related work combines the analysis of functional variability with the exploration of mixed-critical product-lines in the solution space~\cite{Barner2016} as well as the exploration of variants of hardware architecture~\cite{Eder2020}. Furthermore,  ~\cite{Dieudonne2021RMC} presents concepts for reuse of software-component and certification-artefacts in avionics software.

\noindent\textbf{Maturity and gap.} TRL~5-6, open-source and civilian-proven. The asset supports product-line modelling and analysis; the defence gap is to demonstrate this on defence-representative architectures and toolchains.

\subsubsection{Continuous Assurance and Resilient AI}
\label{sec:contrib-assurance}
\noindent\textbf{Concept.} \autoref{sec:continuous-assurance-resilient-ai} named two problems: keeping assurance evidence current as a sub-system or component changes, and making sure the decisions of an AI component can actually be reviewed and trusted.

\noindent\textbf{Why it matters for SDD.} An SDD system that only re-certifies from scratch after every patch cannot keep pace with weekly updates. For AI components, a classification without a justification trail gives a human operator nothing to check before acting on it.

\noindent\textbf{Civilian evidence.} The Evidential Tool Bus~\footnote{https://www.fortiss.org/en/results/software/evidential-toolbus-2} provided a framework for continuous, accountable software certification. The FOCETA\footnote{https://www.fortiss.org/en/research/projects/detail/foceta} project provides formal-methods-based verification and robustness assessment for AI/ML components. KI WISSEN\footnote{https://www.kiwissen.de/} developed hybrid AI methods combining data-driven and knowledge-based approaches for safety-critical automated-driving functions, including explainable, interpretable perception for object and pedestrian recognition.

\noindent\textbf{Maturity and gap.} Both concepts are provided at TRL~5-6, civilian-validated. The Evidential Tool Bus's continuous evidence-generation approach already produces and timestamps an updated assurance record each time an SDD is patched, giving certifying authorities an auditable trail of what changed and why. The explainable justification trail developed in the FOCETA and KI Wissen projects is structurally the same problem as combining sensor-fused threat classification with explicit rule-based engagement constraints, which exactly supports the human-in-the-loop review that meaningful human control requires (\autoref{sec:challenges-ai}).

\section{Recommendations: A Call to Action Towards 2030}
\label{sec:vision}
\begin{figure}[htp!]
  \centering
  \includegraphics[width=\textwidth]{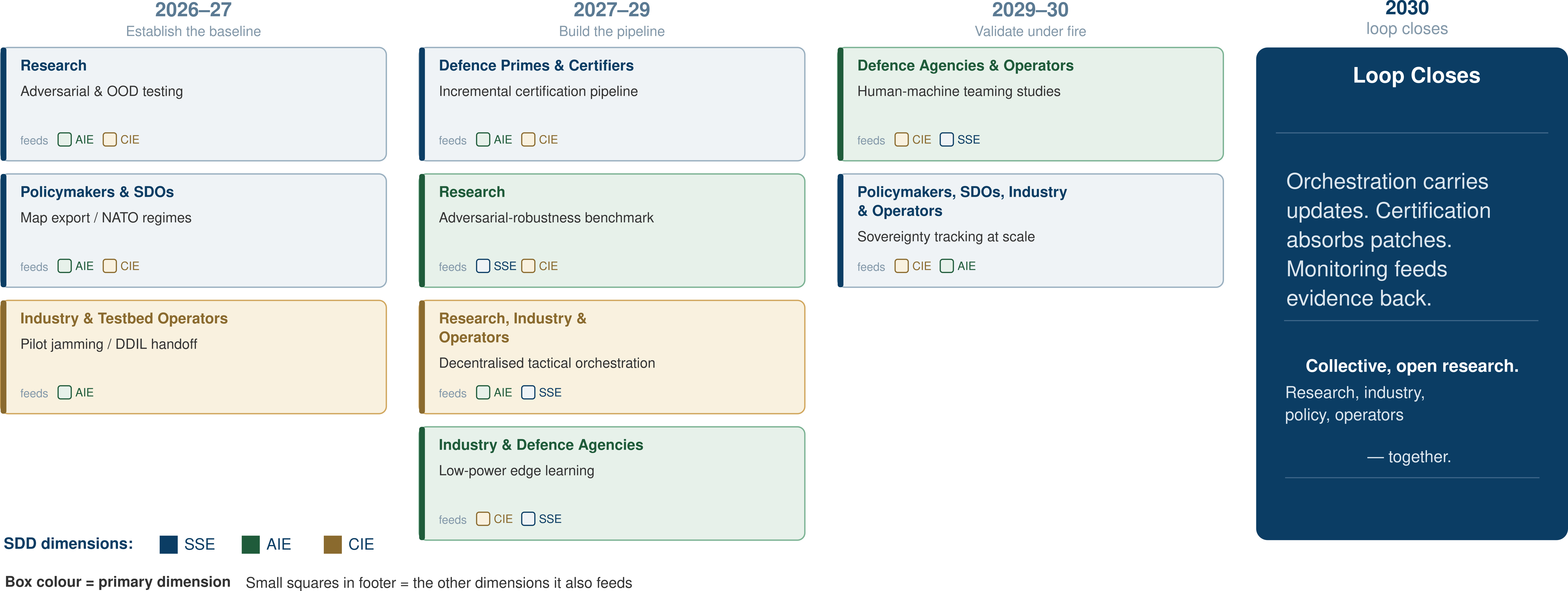}
  \caption{fortiss SDD vision, roadmap to closing the loop until 2030.}
  \label{fig:loop-roadmap}
\end{figure}
 
\autoref{sec:contributions} showed, capability by capability, that methodologically sound, often externally validated, civilian evidence already exists for every stage of the SDD loop in
\autoref{sec:capabilities}, and named the specific gap between that evidence and a defence, adversarial, or certification context. Closing that gap requires researchers willing to redirect existing methods towards adversarial and contested conditions, industry willing to expose civilian tooling to real
operational requirements, policymakers willing to shape the regulatory instruments this transfer depends on, and a public and research community willing to validate and explore/improve the outcome.
This section sets out what each of these actors is best placed to do, organised by the time horizon (short, medium, long term) on which the work becomes possible and by the SDD dimension (SSE, AIE, CIE) it advances. The overall timeframe is presented in \autoref{fig:loop-roadmap}.
 
The time scales proposed are not independent. A certification pipeline (medium term) cannot be demonstrated before adversarial testing produces evidence worth certifying (short term); an open adversarial-robustness benchmark cannot be built before that same testing produces the failure catalogue it would be built from; and human-machine-teaming interfaces (long term) cannot be
validated before the assurance and explainability foundation the medium term establishes is in place. Short-term work builds the baseline; medium-term work builds the pipeline; long-term work closes the loop under conditions that resemble the real thing.

\subsection{Short Term (2026-2027): Testing, Compliance, and DDIL Connectivity Baseline}
\label{sec:vision-short}
 
\noindent\textbf{Research: make adversarial and out-of-distribution testing the default target.} \autoref{sec:testing} and \autoref{sec:contrib-testing} identified the same gap from two angles: search-based testing frameworks such as OpenSBT are validated in nominal-fault and edge-case UAS scenarios, not jamming, DDIL, or sensor denial. Closing this gap is what lets search-based testing feed the certification pipeline with evidence beyond nominal conditions, and it is a matter of redirecting existing, actively developed methods rather than building new tooling from scratch.

The research community already has the building blocks: complete simulator approaches that reduce the Sim2Real gap~\cite{sorokin2026ensembles}, systematic fault injection test case generation~\cite{munaro2025faultinjection}, explainability guided fuzzing for deep learning systems, and diffusion-based domain enhancements for test case generation. The near-term research task is a contained transfer study, run by any group with access to a representative perception or navigation model: apply an existing search-based testing pipeline, unmodified in its search mechanism, with a fitness objective redirected to the failure class under study. Redirected to sensor-denial objectives, the same pipeline stresses a perception model's behaviour under degraded or missing input, the graceful-degradation evidence \autoref{sec:challenges-ai} calls for; redirected to jamming and DDIL objectives, it produces the failure catalogue a short-term connectivity pilot underneath needs to prioritise which degraded-channel conditions to test against.

\noindent\paragraph{Policymakers and SDOs: start mapping defence-specific regulatory regimes now.} \autoref{sec:compliance} and \autoref{sec:contrib-compliance} argued that compliance must run continuously rather than as a milestone audit. Continuous security-compliance research
already spans industrial DevSecOps case studies\cite{moyon2024industrial}, automated security finding management\cite{voggenreiter2024automated}, and automated extraction of change-impact obligations from evolving regulation\cite{angermeir2024towards}. Extending this change-impact methodology to export control and classified dependency tracking is a mapping exercise that defence policy makers are best positioned to trigger now, in parallel with the empirical work on software-supply-chain provenance and SBOM practice already under way in open-source ecosystems, the natural technical basis for the classified dependency and multi-vendor provenance tracking \autoref{sec:challenges-sse} demands. The same continuous-compliance methodology already underpins EU AI Act tooling and early argumentation frameworks for adversarial robustness and regulatory compliance of LLM-based components, so extending it costs little beyond the mapping work itself. NATO interoperability requirements, the third item this mapping exercise must cover, are where SDOs' role stops being paperwork: FMN-compliant architectures are a connectivity standard, not a compliance checkbox, and mapping them now is what lets the short-term connectivity pilot below inherit a standards basis rather than build one from scratch.
 
\noindent\paragraph{Industry and testbed operators: pilot connectivity under representative QoS and jamming constraints.} \autoref{sec:infrastructure} and \autoref{sec:contrib-infrastructure} named the TSN-to-DTN handover,
and its reverse on link restoration, as unsolved. The building blocks already exist. Multi-AP interference mitigation and dynamic resource-unit sharing have already been validated under Wi-Fi~6/7 contention\cite{mada2026exploratoryanalysiswifi6}, a realistic proxy for the degraded-channel conditions a jamming pilot would need. Furthermore, an IESG-approved DetNet controller-plane framework\cite{rfc9938} is available together with comprehensive contributions to IETF, IRTF, and ETSI, which already provides the standardised basis mentioned in \autoref{sec:contrib-infrastructure}. Industry partners with access to a DDIL testbed, together with research institutions holding the underlying softwarized infrastructure assets, are best placed to run a contained pilot proving the handoff under representative constraints. The same testbed fullfills a double duty. Run alongside the perception and navigation models mentioned before, rather than as a connectivity-only exercise, it turns the jamming and sensor-denial scenarios research is searching for into physically realistic test conditions, instead of simulated approximations.

\subsection{Medium Term (2027-2029): Build the Certification, Assurance, Orchestration Pipelines}
\label{sec:vision-medium}
 
\noindent\textbf{Defence primes and certification authorities: demonstrate incremental certification end to end.} The lifecycle paradox (\autoref{sec:lifecycle}) depends on traceable modelling and on continuous evidence-generation approaches reaching a shared,
demonstrated pipeline, rather than the remaining two capabilities that merely coexist. For this to occur, research institutions can supply
the traceable models and evidence-generation tooling, but only an authority with certification standing can validate that the resulting pipeline actually satisfies incremental re-certification in practice. Explainable, interpretable perception work for object and pedestrian recognition is the closest available template for the justification trail such a pipeline must produce for a sensor-fused threat classification, and runtime monitoring of learned-component behaviour is the mechanism that would keep such an assurance
argument current between re-certification events. What keeps that argument current is not the model alone: a monitored component whose updates cannot reach the platform or whose runtime telemetry cannot reach the monitor breaks the assurance chain just as surely as an unvalidated model would, which is why the orchestration work described next is not a parallel workstream but a dependency of certification itself.
 
\noindent\paragraph{Research: build an open adversarial-robustness benchmark from the short-term findings.} Once short-term transfer studies produce real
jamming, DDIL, and sensor-denial-induced failure cases, the medium-term task is turning those cases into a benchmark and methodology for defence-relevant perception and navigation models, extending quantitative coverage analysis and targeted-boundary-testing metrics already used to evaluate ADS test suites developed openly enough, in alignment with the EU sovereign stack principles, and hosted independently enough, to become a reference point across the research community. Certification authorities can cite it directly as the justification-trail evidence the previous paragraph calls for, sparing every subsequent certification effort from re-deriving robustness evidence from scratch; and because the failure cases span connectivity-degraded as well as sensor-degraded conditions, the same benchmark brings double benefit.
 
\noindent\paragraph{Research, Industry and Operators: extend decentralised orchestration from civilian edge-cloud to tactical networks.} \autoref{sec:infrastructure} identified resilience under denied, disrupted, intermittent, and destroyed infrastructure as a first-class SDD requirement, and \autoref{sec:contrib-infrastructure} pointed to store-carry-forward and in-network caching as the relevant civilian evidence. Cognitive, decentralised edge-cloud orchestration as in CODECO~\cite{sofia2026scalablefederatedcontainerorchestration} has already been validated for autonomous mobile systems (robots) at the network edge\cite{zhu2026codeco}, including greenness and energy-aware scheduling in far-edge Kubernetes deployments. Robotic edge orchestration under variable connectivity and energy budgets is a close civilian proxy for a forward-deployed C2 or sensor-fusion node operating without guaranteed backhaul. Emerging pan-European infrastructure such as Euro-3C, the recent federated telco-edge-cloud pilot spanning more than 70 nodes in 13 countries, built on the same open frameworks (Sylva among them) discussed in \autoref{sec:contrib-infrastructure}, is already validating this orchestration model at production scale, albeit for civilian sovereignty rather than tactical resilience objectives. The medium-term task, best undertaken jointly by research institutions and the infrastructure (telecommunications, Cloud) operators who run edge-cloud testbeds, is extending this decentralised orchestration logic to operate across the deterministic, opportunistic, and delay-tolerant regimes of \autoref{sec:infrastructure} jointly, rather than validating orchestration and connectivity resilience as separate problems. What travels across that orchestration layer is not only raw data: a forward-deployed node's on-device inference results and its own re-certification evidence move through the same store-carry-forward paths, so the scheduling decisions this pilot validates are implicitly deciding how much edge-AI compute a denied node can sustain and how long a certification argument can stay current before it goes stale.
 
\noindent\paragraph{Industry and defence agencies: extend low-power, on-device learning to defence-representative platforms.} \autoref{sec:nc}/\autoref{sec:contrib-nc} showed validated neuromorphic edge execution  across wearable, gesture-control, drone-vision, and civilian and dual-use satellite contexts, with recent results giving quantified rather than general efficiency claims: validated energy and latency gains for neuromorphic force control in an industrial task and event-driven, spike-based LiDAR processing for fast and energy-efficient object detection\cite{neumeier-spikeclouds}. The medium-term step requires an industry or defence-agency partner able to provide an actual platform for validation, for instance, a wearable under real size, weight, power and DDIL constraints. Validating under DDIL constraints specifically, rather than benign lab conditions, is what makes this an edge-AI result and not just an efficiency result: a platform that only performs well with guaranteed backhaul has not actually demonstrated the required on-device autonomy \autoref{sec:nc}, and every validation run is itself the kind of platform-level update that the certification pipeline above will eventually need to absorb.
 
\subsection{Long Term (2029-2030): Validate the Loop Under Realistic Operational Conditions}
\label{sec:vision-long}
 
\noindent\textbf{Defence agencies and operators: validate human-machine teaming through operator studies.} Meaningful human control (\autoref{sec:challenges-ai}) is a legal and engineering
requirement, not yet a validated interface. The long-term target is concrete override, intervention, and abort mechanisms, tested with actual operators under representative mission conditions, building on the explainability foundation and the LLM-adversarial-robustness assurance argumentation named in \autoref{sec:vision-medium}: an operator cannot meaningfully override a decision they cannot inspect, and that inspection interface is what the medium term's assurance work is meant to feed. Only defence agencies with access to real operators, in structured studies designed
with human factor researchers, can close this validation gap. Two conditions have to hold before that gap can be tested honestly: the override command must reach the platform under the connectivity regime actually in operation or the study tests an interface that will not survive contact with a denied link. 
 
\noindent\paragraph{Policymakers, SDOs, Industry, Operators: track sovereignty obligations across a full defence programme.} Compliance-by-design (\autoref{sec:compliance}/\autoref{sec:contrib-compliance}) and variability and reuse management (\autoref{sec:variability}/\autoref{sec:contrib-variability}) converge here: SBOM obligations, export-control status and multinational interoperability requirements including, e.g., AI model versions on that same SBOM ledger must be tracked jointly across a representative defence software stack, which only a programme owner with authority over the full stack can do at scale. Multinational interoperability is where this stops being a paperwork exercise: the deterministic-to-opportunistic handover proven within one country's testbed does not by itself survive a multinational deployment. So tracking these obligations at programme scale is inseparable from securing the cross-border spectrum access and testbed-sharing agreements the handover needs at alliance scale.
\newpage
\footnotesize
\bibliographystyle{IEEEtranDOI}
\bibliography{sdd-references}
\section{Referenced Web Sources}
\label{app:urls}
\small
The following web sources are cited as footnotes throughout this paper, listed in order of first appearance.
\footnotesize
\begin{enumerate}
  \item German National Academy of Science and Engineering (acatech) — \href{https://en.acatech.de/}{en.acatech.de}
  \item OWASP Top 10 (2025) — \href{https://owasp.org/Top10/2025}{owasp.org/Top10/2025}
  \item Corvus Intelligence, European defence tech market 2025 — \href{https://corvusintell.com/blog/defense-market/defense-tech-market-europe-2025/}{corvusintell.com/blog/defense-tech-market-europe-2025}
  \item Council on Foreign Relations, NATO 2\% spending milestone — \href{https://www.cfr.org/expert-brief/nato-countries-reach-spending-milestone-2-percent-enough}{cfr.org/expert-brief/nato-2-percent-spending}
  \item German Federal Ministry of Defence (BMVg), Bundeswehr special fund (Sondervermögen) — \href{https://www.bmvg.de/en/news/over-eur-100-billion-for-the-bundeswehr-and-for-our-security-5362626}{bmvg.de/en/news/over-eur-100-billion-bundeswehr}
  \item Bundeswehr, D-LBO digitalisation programme — \href{https://www.bundeswehr.de/de/meldungen/digitalisierung-landbasierte-operationen}{bundeswehr.de/de/meldungen/digitalisierung-landbasierte-operationen}
  \item Arquus, French SCORPION programme — \href{https://www.arquus-defense.com/scorpion-program}{arquus-defense.com/scorpion-program}
  \item European Commission, SAFE loan instrument — \href{https://commission.europa.eu/topics/defence/future-european-defence}{commission.europa.eu/topics/defence/future-european-defence}
  \item European Commission, AGILE fast-fielding fund — \href{https://commission.europa.eu/priorities-2024-2029/security-and-defence}{commission.europa.eu/priorities-2024-2029/security-and-defence}
  \item NATO, Revised Artificial Intelligence Strategy (2024) — \href{https://www.nato.int/en/about-us/official-texts-and-resources/official-texts/2024/07/10/summary-of-natos-revised-artificial-intelligence-ai-strategy}{nato.int, Revised AI Strategy (2024)}
  \item EURO-3C, federated Telco-Edge-Cloud (3C) infrastructure project — \href{https://euro-3c.eu/}{euro-3c.eu}
  \item Fraunhofer, multi-agent reinforcement learning and autonomous cooperation — \href{https://safe-intelligence.fraunhofer.de/en/articles/multi-agent-reinforcement-learning-when-intelligent-systems-must-cooperate-autonomously}{safe-intelligence.fraunhofer.de, multi-agent RL article}
  \item SeRANIS, B5G/6G laboratory environment — \href{https://seranis.de/}{seranis.de}
  \item dtec.bw, SeRANIS project page — \href{https://dtecbw.de/home/forschung/unibw-m/projekt-seranis}{dtecbw.de, SeRANIS project page}
  \item Fraunhofer IKS, Software-Defined Defense — \href{https://www.iks.fraunhofer.de/en/topics/software-defined-defense.html}{iks.fraunhofer.de/en/topics/software-defined-defense}
  \item Sylva Project, open-source telco cloud stack — \href{https://sylvaproject.org/}{sylvaproject.org}
  \item CODECO, cognitive decentralised edge-cloud orchestration — \href{https://he-codeco.eu/}{he-codeco.eu}
  \item Object Management Group (OMG), Data Distribution Service (DDS) portal — \href{https://www.omg.org/omg-dds-portal/}{omg.org/omg-dds-portal}
  \item DARPA, SyNAPSE programme — \href{https://www.darpa.mil/research/programs/systems-of-neuromorphic-adaptive-plastic-scalable-electronics}{darpa.mil, SyNAPSE programme}
  \item fortiss, AutoFOCUS3 (AF3) MBSE platform — \href{https://af3.fortiss.org}{af3.fortiss.org}
  \item NASA/JPL, Interplanetary Overlay Network — \href{https://www.nasa.gov/directorates/somd/space-communications-navigation-program/interplanetary-overlay-network/}{nasa.gov, Interplanetary Overlay Network}
  \item Intel, Loihi 2 neuromorphic computing technology brief — \href{https://www.intel.com/content/www/us/en/research/neuromorphic-computing-loihi-2-technology-brief.html}{intel.com, Loihi 2 technology brief}
  \item DARPA, Guaranteeing AI Robustness against Deception (GARD) programme — \href{https://www.darpa.mil/research/programs/guaranteeing-ai-robustness-against-deception}{darpa.mil, GARD programme}
  \item EU AI Act Compliance Checker — \href{https://artificialintelligenceact.eu/assessment/eu-ai-act-compliance-checker/}{artificialintelligenceact.eu, Compliance Checker}
  \item fortiss, Evidential Tool Bus — \href{https://www.fortiss.org/en/results/software/evidential-toolbus-2}{fortiss.org, Evidential Tool Bus}
  \item fortiss, FOCETA project — \href{https://www.fortiss.org/en/research/projects/detail/foceta}{fortiss.org, FOCETA project}
  \item KI Wissen project — \href{https://www.kiwissen.de/}{kiwissen.de}
\end{enumerate}

\appendix
\section{Research Infrastructures Referenced in this Paper}
\label{app:labs}
\begin{itemize}
  \item \textbf{fortiss Mobility Lab} (AutoFOCUS3, architecture exploration, OpenSBT) — \url{https://www.fortiss.org/forschung/fortiss-labs/detail/mobility-lab}
  \item \textbf{fortiss Neuromorphics Lab} (EMMANÜELA, event-based sensing) — \url{https://www.fortiss.org/en/research/fortiss-labs/detail/neuromorphics-lab}
  \item \textbf{fortiss IIoT Lab} (TSNWiFi, DetNetWiFi, DABBER, IETF DetNet/RAW) — \url{https://www.fortiss.org/en/research/fortiss-labs/detail/iiot-lab}
\end{itemize}

\end{document}